\documentclass[reprint,
superscriptaddress,
 amsmath,amssymb,
 aps,
prx,
]{revtex4-1}

\usepackage[utf8]{inputenc}
\usepackage[T1]{fontenc}
\usepackage{braket}
\usepackage{graphicx}
\usepackage{subcaption}
\usepackage{mwe}
\usepackage{scalerel}
\usepackage{amsmath}
\usepackage{dsfont}
\usepackage{xr}

\begin{document}
\title{Interaction of atomic systems with quantum vacuum beyond electric dipole approximation}

\author{Miriam Kosik}
\affiliation{Institute of Physics, Faculty of Physics, Astronomy and Informatics, Nicolaus Copernicus University, Grudziadzka 5, 87-100 Torun, Poland}

\author{Oleksandr Burlayenko}
\affiliation{Department of Physics and Technology, V.N. Karazin Kharkiv National University, Kharkiv,
Ukraine}

\author{Ivan Fernandez-Corbaton}
\affiliation{Institute of Nanotechnology, Karlsruhe Institute of Technology, 76021 Karlsruhe, Germany}

\author{Carsten Rockstuhl}
\affiliation{Institute of Theoretical Solid State Physics, Karlsruhe Institute of Technology, 76131 Karlsruhe, Germany}
\affiliation{Institute of Nanotechnology, Karlsruhe Institute of Technology, 76021 Karlsruhe, Germany}

\author{Karolina S\l{}owik}
\affiliation{Institute of Physics, Faculty of Physics, Astronomy and Informatics, Nicolaus Copernicus University, Grudziadzka 5, 87-100 Torun, Poland}

\email{mkosik@doktorant.umk.pl, karolina@fizyka.umk.pl}


\begin{abstract}
The photonic environment can significantly influence emission properties and interactions among atomic systems. In such scenarios, frequently the electric dipole approximation is assumed that is justified as long as the spatial extent of the atomic system is negligible compared to the spatial variations of the field. While this holds true for many canonical systems, it ceases to be applicable for more contemporary nanophotonic structures. To go beyond the electric dipole approximation, we propose and develop in this article   an analytical framework to describe the impact of the photonic environment on emission and interaction properties of atomic systems beyond the electric dipole approximation. Particularly, we retain explicitly magnetic dipolar and electric quadrupolar contributions to the light-matter interactions. We exploit a field quantization scheme based on electromagnetic Green's tensors, suited for dispersive materials. We obtain expressions for spontaneous emission rate, Lamb shift, multipole-multipole shift and superradiance rate, all being modified with dispersive environment. The considered influence could be substantial for suitably tailored nanostructured photonic environments, as demonstrated exemplarily.
\end{abstract}

\flushbottom
\maketitle
\thispagestyle{empty}
\section*{Introduction}

\noindent An excited atomic system, e.g. an atom, a molecule, a quantum dot, can decay radiatively from an excited to a ground state while releasing its energy into
the photonic environment. The rate of this process depends on the properties of the environment, and in a pioneering work by Purcell the
possibility to control the spontaneous emission lifetimes of atomic systems by tailoring their surroundings
was first investigated\cite{purcell1946}. Hence, it has also been called the Purcell-effect. The effect has been experimentally verified in various types of cavities or band-gap environments, including semiconductor microstructures \cite{Gerard1998}, photonic crystals \cite{Englund2005}, and plasmonic nanoparticles, where the emission rate was enhanced up to three orders of magnitude \cite{Yuan2013,Akselrod2014}. Besides emission enhancement on its own, photonic environment equally affects the interactions between multiple atomic systems. In vacuum, examples such as dipole-dipole coupling \cite{Ficek1987}, or collective phenomena like Dicke superradiance have been explored \cite{Dicke1954,Gross1982}. Also, all these effects can be tailored by suitably engineering the photonic environment \cite{Dzsotjan2010, Bouchet2019}.

The influence of photonic environment on these phenomena is usually quantified in electric dipole approximation. This is justified when the electric field shows a negligible spatial variation across the size of the atomic system. Steps beyond may be required if the atomic system is large with respect to wavelength of light it is coupled to\cite{Konzelmann2019} or, contrary, if the electromagnetic field is focused into spots comparable in size to the atomic systems. The latter can be realized by nanoscopic environments and picocavities, capable to localize the electric field into nanometric spatial domains, providing high intensities and spatial modulations at the length scale of tens of nanometers. This tends to be comparable to the size-scale of molecules or quantum dots \cite{Schuller2010,Gramotnev2010, Benz2016}. Then the usual mismatch of size-scales of photonic modes and atomic systems is reduced, which leads to enhanced interaction probability if the photonic modes and the atomic system overlap in space. 
Furthermore and potentially more interestingly, nanostructured environments may also open light-matter interaction channels beyond that one corresponding to coupling of electric field with electric dipoles. For example, light concentrations in nanoscopic regions implies modulations of electromagnetic field at spatial distances comparable to the size of atomic systems, which may couple to electric quadrupolar or higher-order moments. In addition, due to their large refractive index, dielectric nanomaterials offer the possibility of strong concentrations of magnetic fields \cite{Schmidt2012}. This prompts to consider not just electric multipolar contributions but at the same time their magnetic counterparts. Until now, enhancement of the rate of a magnetic dipole emission by a nanostructure was considered \cite{Hein2013} and reported experimentally in lanthanide ions \cite{Karaveli2010,Taminiau2012,Kasperczyk2015,Vaskin2018}. 
Large enhancement of quadrupole \cite{Kern2012,Filter2012,Yannopapas2015} and even higher-order transitions was equally predicted \cite{Rivera2016,Neuman2018}. Transitions driven with several multipolar mechanisms have been considered \cite{Tighineanu2014} and observed \cite{Li2018} respectively in semiconductor quantum dots and transition-metal/lanthanide ions. Recently, it was pointed out that the simultaneous existence of several interaction channels enhanced in intensity with photonic nanostructures might open new avenues on the route to control spontaneous emission, related to their controlled interference \cite{Rusak2019}. 
In that work, however, a semiclassical scheme was proposed to account for interference of different multipolar spontaneous emission mechanisms based on \textit{a priori} knowledge on transition rates through individual channels. To go beyond this semiclassical treatment, we develop here an \textit{ab initio} analytic theory based on field quantization in dispersive media, to evaluate not just transition rates, but also energy shifts in the cases of individual or multiple emitters. We derive expressions for spontaneous emission rate, Lamb shift, collective emission rates, and interaction strengths of atomic systems in structured dispersive environments, while considering the magnetic dipolar and electric quadrupolar interaction channels. The optical properties of the environment are expressed in terms of electromagnetic Green’s tensor, defined by the spatial and spectral dependence of the electric permittivity, that defines the photonic environment. Our result is an extension to those results previously obtained in electric-dipole approximation in Refs.~\cite{Barnett1996,Dzsotjan2010}. Here, we include two next-order terms of the multipolar coupling Hamiltonian~\cite{barron}, namely the magnetic dipole and the electric quadrupole. This unlocks qualitatively new routes to tailor light-matter coupling exploiting interference effects.

The work is organized as follows: in the next section we introduce the general framework used to describe light-matter coupling in the presence of dispersive and absorptive structured materials beyond electric-dipole approximation. Examples of application of the theory to specific geometries are given in Section \ref{sec:example} with detailed numerical results provided in the Supplementary Material. Lengthy calculations are documented in Appendices. 

\section{Theory}
In this section we derive expressions for spontaneous emission rate and Lamb shift of a single atomic system in such environment at first. Next, we generalize these expressions to many-atom systems. 

\subsection{\label{subsec:atom}Atomic system}
We assume that the atomic system can be approximated by two active energy levels, separated by energy $\hbar\omega_\mathrm{0}$, where $\hbar$ stands for the reduced Planck's constant. The corresponding ground and excited states are denoted by $\ket{g}$ and $\ket{e}$, respectively. The system is fully described by a set of Pauli operators: the lowering operator $\sigma = \ket{g}\bra{e}$ and the inversion operator $\sigma_z = \ket{e}\bra{e}-\ket{g}\bra{g}$, following the usual commutation rules $[\sigma,\sigma^\dagger]= -\sigma_z$, $[\sigma,\sigma_z]=2\sigma$. The free Hamiltonian of the system reads $\mathcal{H}_0=\hbar\omega_0\sigma^\dagger\sigma$. This Hamiltonian can be generalized to the case of multiple emitters in a straightforward manner, as done in one of the following sub-sections.

\subsection{\label{subsec:field}Quantized electromagnetic field in dispersive media}
In this work we follow the quantization scheme in dispersive and absorbing media, developed in Refs.~\cite{Huttner1992,Matloob1995,Gruner1996,Dung1998}. We restrict ourselves to nonmagnetic matter with relative permeability $\mu=1$. The constitutive equation relating the temporal Fourier components of the displacement field $\mathbf{D}(\mathbf{r},\omega)$, the electric field $\mathbf{E}(\mathbf{r},\omega)$, and medium polarization $\mathbf{P}(\mathbf{r},\omega)$ in an absorbing medium takes the form $\mathbf{D}(\mathbf{r},\omega)=\epsilon_0 \mathbf{E}(\mathbf{r},\omega)+\mathbf{P}(\mathbf{r},\omega) = \epsilon_0 \epsilon(\mathbf{r},\omega) \mathbf{E}(\mathbf{r},\omega)+\mathbf{P}_N(\mathbf{r},\omega)$. Here, $\mathbf{P}_N(\mathbf{r},\omega)$ describes a noise contribution to the polarization $\mathbf{P}(\mathbf{r},\omega)$ arising from vacuum fluctuations in an absorbing medium. The related noise current density reads $\mathbf{j}_N(\mathbf{r},\omega) = -i\omega\mathbf{P}_N(\mathbf{r},\omega)$. 

Since we investigate vacuum-induced effects, we are interested in the case of a vanishing mean electric field. Then, the only field is related to noise current fluctuations and can be expressed as 
\begin{equation}
\mathbf{E}\left(\mathbf{r},\omega\right) = i\mu_0\omega\int d^3 r^\prime \mathbf{G}\left(\mathbf{r},\mathbf{r}^\prime,\omega\right)\mathbf{j}_N\left(\mathbf{r}^\prime,\omega\right),
\end{equation}
where $\mu_0$ stands for vacuum permeability.

The dyadic tensor $\mathbf{G}\left(\mathbf{r},\mathbf{r}^\prime,\omega\right)$ is the full electromagnetic Green's tensor characterizing the environment, determined from the Maxwell-Helmholtz equation
\begin{equation}
\left[\nabla\times\nabla\times
-\frac{\omega^2}{c^2}\epsilon\left(\mathbf{r},\omega\right)\right]\mathbf{G}\left(\mathbf{r},\mathbf{r}^\prime,\omega\right)=I\delta\left(\mathbf{r}-\mathbf{r}^\prime\right),
\end{equation}
with $I$ representing the unit dyadic. The boundary condition for the Green's tensor at infinity reads \mbox{$\mathbf{G}\left(\mathbf{r},\mathbf{r}^\prime,\omega\right)\rightarrow 0$} for \mbox{$\left|\mathbf{r}-\mathbf{r}^\prime\right|\rightarrow \infty$}. The Green's tensor serves as the kernel that connects the electric field $\mathbf{E}\left(\mathbf{r},\omega\right)$ with its source at position $\mathbf{r}^\prime$.

The requirement for the canonical equal-time commutation relations of quantized fields to hold allows one to express the noise current in the form \cite{Huttner1992,Matloob1995,Gruner1996}
\begin{equation}
\mathbf{j}_N(\mathbf{r},\omega)=\omega\sqrt{\frac{\hbar\epsilon_0}{\pi}\mathrm{Im}\epsilon(\mathbf{r},\omega)}\,\mathbf{f}_\omega\left(\mathbf{r}\right).
\end{equation}
Here, $\epsilon_0$ represents the electric permittivity of vacuum and $\epsilon\left(\mathbf{r},\omega\right) = \mathrm{Re}\,\epsilon \left(\mathbf{r},\omega\right) + i\, \mathrm{Im}\,\epsilon \left(\mathbf{r},\omega\right)$ is the relative permittivity of the dispersive and absorptive medium surrounding the atomic system. For simplicity, we assume isotropic media, so that the permittivity can be expressed as a scalar function. The bosonic operator fields take the form \mbox{$\mathbf{f}_\omega\left(\mathbf{r}\right)=\sum_\mathrm{j}f_{\omega,\mathrm{j}}\left(\mathbf{r}\right)\mathbf{e}_\mathrm{j}$}, where $\mathrm{j}\in\{x,y,z\}$ and $\mathbf{e}_\mathrm{j}$ is a unit vector in the $\mathrm{j\textsuperscript{th}}$ direction.
They obey the following commutation relations \cite{Huttner1992,Matloob1995,Gruner1996}
\begin{eqnarray}\label{eq:f_commutators}
\left[f_{\omega,\mathrm{j}}\left(\mathbf{r}\right),f_{\omega^\prime,\mathrm{k}}\left(\mathbf{r}^\prime\right)\right]&=&0,\\
\left[f_{\omega,\mathrm{j}}\left(\mathbf{r}\right),f^\dagger_{\omega^\prime,\mathrm{k}}\left(\mathbf{r}^\prime\right)\right]&=&\delta_\mathrm{jk}\delta(\omega-\omega^\prime)\delta(\mathbf{r}-\mathbf{r}^\prime).\nonumber
\end{eqnarray}
Finally, the electric field can be expressed in terms of bosonic operators as follows
\begin{equation}
\mathbf{E}\left(\mathbf{r},\omega\right) = i\sqrt{\frac{\hbar}{\pi\epsilon_0}}\frac{\omega^2}{c^2}\int d^3 r^\prime \sqrt{\mathrm{Im}\,\epsilon\left(\mathbf{r}^\prime,\omega\right)}\,\mathbf{G}\left(\mathbf{r},\mathbf{r}^\prime,\omega\right)\mathbf{f}_\omega\left(\mathbf{r}^\prime\right),
\end{equation}
where $c$ is the vacuum speed of light.

In the Schr\"{o}dinger picture, the positive frequency part $\mathbf{E}^{(+)}\left(\mathbf{r}\right)=\left[\mathbf{E}^{(-)}\left(\mathbf{r}\right)\right]^\dagger$ of the electric field operator $\mathbf{E}\left(\mathbf{r}\right)=\mathbf{E}^{(+)}\left(\mathbf{r}\right)+\mathbf{E}^{(-)}\left(\mathbf{r}\right)$ is obtained through the integration
$\mathbf{E}^{(+)}\left(\mathbf{r}\right) = \int_0^\infty d\omega \, \mathbf{E}\left(\mathbf{r},\omega\right)$.
Similarly, the positive frequency part of the magnetic field is expressed given by
$\mathbf{B}^{(+)}\left(\mathbf{r}\right) = \int_0^\infty d\omega \, \mathbf{B}\left(\mathbf{r},\omega\right)$,
where the frequency components of the magnetic fields are connected to the corresponding electric field through 
$\mathbf{B}\left(\mathbf{r},\omega\right)= -\frac{i}{\omega}\nabla\times\mathbf{E}\left(\mathbf{r},\omega\right)$.

The free-field Hamiltonian reads 
\mbox{$\mathcal{H}_\mathrm{field} = \int d^3 r \int d\omega \,\, \hbar\omega \,\, \mathbf{f}_\omega\left(\mathbf{r}\right)^\dagger \mathbf{f}_\omega\left(\mathbf{r}\right)$.}

\subsection{\label{subsec:coupling}Coupling Hamiltonian}
The electric dipole approximation is based on the assumption that the electric field can be approximated as constant across the spatial extent of the atomic system. In particular, this means that direct coupling of the emitter with the magnetic field is ignored and spatial modulations of the electric fields are neglected as well. 
Typically, the above assumption is very well met: the scale of spatial modulations of the electric field is set by its wavelength, usually in the optical or near-infrared range, while the modulations of the field's envelope are even slower and thus negligible. This assumption holds true in free space or traditional cavities, if the atomic system is represented by an atom or a molecule. However, the assumption might no longer be applicable if the emitter is positioned within a subwavelength electromagnetic hotspot, e.g. in photonic crystal cavities, close vicinity to plasmonic nanostructures, near picocavities or even in free space when geometrically large emitters like semiconductor quantum dots are considered \cite{Tighineanu2014,Neuman2018}. 
For this reason, we include two higher-order terms of the multipolar coupling Hamiltonian, which include first-order spatial derivatives of the electric field~\cite{barron}. The Hamiltonian is given in the rotating wave approximation, valid as long as the coupling strengths are small with respect to the transition frequency $\omega_0$
\begin{widetext}
\begin{equation}\label{eq:hamiltonian}
\mathcal{H}_\mathrm{int} = -\left[
\mathbf{E}^{(-)}_0\cdot\mathbf{d} + 
\mathbf{B}^{(-)}_0\cdot\mathbf{m} + 
\nabla\mathbf{E}^{(-)}_0:\mathbf{Q}\right]\sigma 
-\sigma^\dagger\left[
\mathbf{d}^\dagger\cdot\mathbf{E}^{(+)}_0 + 
\mathbf{m}^\dagger\cdot\mathbf{B}^{(+)}_0+ 
\mathbf{Q}^\dagger:\nabla\mathbf{E}^{(+)}_0\right],
\end{equation}
\end{widetext}
where the fields are evaluated at the position of the atomic system $\mathbf{r}_0$, and for brevity we denote \mbox{$\mathbf{E}^{(\pm)}_0\equiv \mathbf{E}^{(\pm)}(\mathbf{r}_0)$}. 
We assume that the electric field derivatives exist at this position, i.e. the atomic system should not be placed directly at the interface between two different media. 
Above, $\mathbf{d}=\bra{g}\hat{\mathbf{d}}\ket{e}$ and $\mathbf{m}
=\bra{g}\hat{\mathbf{m}}\ket{e}$ are the electric and magnetic transition dipole moment elements, and $\mathbf{Q}=\bra{g}\hat{\mathbf{Q}}\ket{e}$ is the electric transition quadrupole moment element, respectively. The dot denotes the standard scalar product of vectors, the double dot product of tensors is defined as $\mathbf{C}:\mathbf{D}\equiv\sum_{ij}C_{ij}D_{ji}$, while $\nabla\mathbf{E}^{(\pm)}$ is a dyadic product. In the case of a real electric quadrupole moment tensor, the quadrupolar contribution to the coupling Hamiltonian can be equivalently rewritten as $\mathbf{Q}^\dagger:\nabla\mathbf{E}^{(+)}\left(\mathbf{r}\right)=\left(\mathbf{Q}^\dagger\mathbf{\nabla}\right)\cdot \mathbf{E}^{(+)}\left(\mathbf{r}\right)=\sum_{ij}Q^\star_{ij}\partial_j E_i^{(+)}\left(\mathbf{r}\right)$. Please note that different degrees of freedom in the operators above are denoted as follows: 
\begin{itemize}
    \item the degree of freedom related to the two-dimensional Hilbert space spanned by $\{\ket{g},\ket{e}\}$ is already included in the symbols of transition moments, and is relevant for Hermitian conjugation, e.g. $\mathbf{Q}^\dagger=(\bra{g}\hat{\mathbf{Q}}\ket{e})^\dagger=\bra{e}\hat{\mathbf{Q}}\ket{g}$; permanent multipole moments are assumed negligible, e.g. $\bra{g}\hat{\mathbf{Q}}\ket{g}=\bra{e}\hat{\mathbf{Q}}\ket{e}=0$,
    \item the dipole moment vectors and the quadrupole moment tensor have elements corresponding to the $x,y,z$ spatial directions, e.g. $d_i$, $Q_{ij}$, responsible for the orientation of the multipolar moment moment,
    \item finally, the fields depend on position in space $\mathbf{r}$, and each element of the Green's tensor is a function of the observation point $\mathbf{r}$ and the source point $\mathbf{r}^\prime$.
\end{itemize}

\subsection{\label{sec:rate}Emission properties of single atomic system}
To derive the spontaneous emission rate of an atomic system, we proceed as follows: First, Heisenberg equations are found for the atomic and the field operators. The equations for the field are then formally integrated, so that the field operators are expressed through the atomic ones. The result is then inserted into atomic equations, so that field variables are completely eliminated from the description. The resulting complicated integro-differential equations are simplified in the Markovian approximation, where the memory effects are neglected. 
As a result, we obtain effective dynamics of the atomic system alone. The procedure is a generalization of the one introduced in Ref.~\cite{Dzsotjan2010}, where only the electric dipole interaction term was taken into account. Since the equations tend to be lengthy, we describe the consecutive steps in detail in Appendix \ref{app:single_system}. The effective evolution of the atomic system reads
\begin{eqnarray}
\label{eq:sigma_dynamics}
\dot{\sigma} &=& -\left[\frac{\gamma}{2}+i\left(\omega_0+\delta\right)\right]\sigma\\
&&-\frac{i}{\hbar}\sigma_z\left[\mathbf{d}^\dagger\cdot\mathbf{E}_{0,\mathrm{free}}^{(+)}+\mathbf{m}^\dagger\cdot\mathbf{B}_{0,\mathrm{free}}^{(+)}+\mathbf{Q}^\dagger:\nabla\mathbf{E}_{0,\mathrm{free}}^{(+)}\right]\nonumber\\
\label{eq:sigmaz_dynamics}
\dot{\sigma}_z &=& -\gamma\left(\sigma_z+\mathds{1}\right)\\
&&+\frac{2i}{\hbar}\sigma^\dagger\left[\mathbf{d}^\dagger\cdot\mathbf{E}_{0,\mathrm{free}}^{(+)}+\mathbf{m}^\dagger\cdot\mathbf{B}_{0,\mathrm{free}}^{(+)}+\mathbf{Q}^\dagger:\nabla\mathbf{E}_{0,\mathrm{free}}^{(+)}\right]\nonumber\\ 
&&-\frac{2i}{\hbar}\left[\mathbf{E}_{0,\mathrm{free}}^{(-)}\cdot\mathbf{d}+\mathbf{B}_{0,\mathrm{free}}^{(-)}\cdot\mathbf{m}+\nabla\mathbf{E}_{0,\mathrm{free}}^{(-)}:\mathbf{Q}\right]\sigma \nonumber,
\end{eqnarray}
where the fields are always evaluated at $\mathbf{r}_0$, $\mathds{1}$ is the identity operator in the atomic Hilbert space, $\gamma$ is the spontaneous emission rate, and $\delta$ stands for the analogue of Lamb shift, calculated beyond the electric dipole approximation. Explicit expressions for these quantities are given in the following. The "free" subscript corresponds to free fields, i.e. fields that are not influenced by atomic back-action. In the vacuum state, these fields account for vacuum fluctuations and their mean value vanishes. The influence of the photonic environment is taken into account through the modified Green's tensor. The emission rate $\gamma$ includes contributions from the electric dipole, magnetic dipole, and electric quadrupole channels, as well as their interference. The rate is expressed as
\begin{equation}\label{eq:gamma_main}
\gamma = \frac{2}{\hbar\epsilon_{0}}\frac{\omega_{0}^{2}}{c^{2}}\sum_{mn}D_m^{r\dagger} D_n^{r^\prime} \mathrm{Im}\,G_{mn}\left(\mathbf{r},\mathbf{r}^{\prime},\omega_0\right)|_{\substack{\mathbf{r}=\mathbf{r_0}
\\{\mathbf{r}^\prime}=\mathbf{r_0}}},
\end{equation}
where we have defined a "generalized transition moment" $\mathbf{D}^r$ with components
\begin{equation}\label{eq:generalized_moment}
D^{r}_m = d_m+\sum_k \left(Q_{mk}+\frac{i}{\omega_0}\sum_p\epsilon_{pkm}m_p\right) \frac{\partial}{\partial r_k},
\end{equation}
with $m,k,p\in \{x,y,z\}$,
$G_{mn}^{\prime\prime}$ denotes the imaginary part of an $mn$ element of the Green's tensor and $\epsilon_{pkm}$ is the Levi-Civita antisymmetric symbol. The derivatives in Eq.~(\ref{eq:generalized_moment}) should be evaluated at the position of the atomic system.
Please note that the "generalized moment" is in fact a differential operator that acts on the Green's tensor, i.e. the "moment" combines atomic and field properties. We stress that $\mathbf{D}^r$ becomes a purely atomic quantity only in the electric dipole approximation. Please note that in this case expression (\ref{eq:gamma_main}) is reduced to the well known form $\gamma = \frac{2}{\hbar\epsilon_{0}}\frac{\omega_{0}^{2}}{c^{2}}\mathbf{d}^\dagger \cdot \mathrm{Im}\,\mathbf{G}\left(\mathbf{r}_0,\mathbf{r}_0,\omega_0\right)\cdot\mathbf{d}$ \cite{Barnett1996,Dzsotjan2010,novotny}.\par 

The Lamb shift in Eq.~(\ref{eq:sigma_dynamics}) is expressed through a principal-value integral:
\begin{eqnarray}\label{eq:lamb_shift}
&&\delta =\frac{1}{\hbar\pi\epsilon_{0}c^{2}}\times\\
&&\times\mathcal{P}\int_0^\infty d\omega\frac{\omega^{2}}{\omega-\omega_{0}}\sum_{mn}D_m^{r\dagger}(\omega) D_n^{r^\prime}(\omega)
\mathrm{Im}\,G_{mn}\left(\mathbf{r},\mathbf{r}^{\prime},\omega\right)|_{\substack{\mathbf{r}=\mathbf{r_0}
\\{\mathbf{r}^\prime}=\mathbf{r_0}}}\nonumber
\end{eqnarray}
where 
\begin{equation}\label{eq:generalized_moment_function}
D^{r}_m\left(\omega\right) = d_m+\sum_k \left(Q_{mk}+\frac{i}{\omega}\sum_p\epsilon_{pkm}m_p\right) \frac{\partial}{\partial r_k},
\end{equation}
and the generalized moment in Eq.~(\ref{eq:generalized_moment}) is \mbox{$D^{r}_m=D^{r}_m\left(\omega_0\right)$}.
Again, in electric dipole approximation expression (\ref{eq:lamb_shift}) reduces to the familiar form derived in Refs.\cite{Dung2000, Dzsotjan2010}. 

\subsubsection*{General comments} Before we move to the next section, we will make a few general comments. 
The Green's tensor $\mathbf{G}\left(\mathbf{r},\mathbf{r}^\prime,\omega\right) = \mathbf{G}_h\left(\mathbf{r},\mathbf{r}^\prime,\omega\right)+\mathbf{G}_s\left(\mathbf{r},\mathbf{r}^\prime,\omega\right)$ can be decomposed into a sum of a homogeneous term $\mathbf{G}_h$ and a scattered part $\mathbf{G}_s$ \cite{Martin1998}. The homogeneous term corresponds to the response in free space or a homogeneous medium, while the scattered one describes influence of scatterers in the environment, e.g. extended interfaces among different media, nanostructured particles, or photonic crystals. The contribution of the homogeneous part of the Green's tensor is already included in the homogeneous-medium spontaneous emission rate or the respective Lamb shift, while the contribution of the scattered part is of general interest. The scattered contribution is frequently finite at the origin, as $\mathbf{r},\mathbf{r}^\prime\rightarrow \mathbf{r}_0$. 

From an analysis of the generalized multipole moment, we find that the electric dipole component depends on the imaginary part of Green's tensor, while the electric quadrupole and the magnetic dipole components are proportional to the sum and difference of the corresponding derivatives of the imaginary part of the Green's function: $Q_{km}\left(\partial_k+\partial_m\right)$, $m_p\left(\partial_k-\partial_m\right)$, $p\neq k,m$, respectively. A simple $\frac{\pi}{4}$-rotation of the coordinate system shows that these can be independently tailored, since there is no general restrictions on the ratios of values of function and its derivatives in different directions at a given point. This observation is a starting point to consider their full interference and engineer environments which might support it \cite{Rusak2019}. 

Generalizing the expressions for the transition rate beyond the electric dipole approximation not only allows to consider corrections to the atomic systems' dynamics in cavities of extreme geometries. More importantly, it is a tool to describe, e.g. optical activity of chiral molecules for which an interplay between the electric and magnetic dipolar coupling of matter and light, i.e. interference of the two transition mechanisms, plays the crucial role. In centrosymmetric systems, parity is a good quantum number, allowing one to identify transitions either described by the electric dipole mechanism or a combination of electric quadrupole and magnetic dipole. Such systems could be considered sources of photons with well-defined parity corresponding to a given transition mechanism. 

\subsection{\label{sec:many-body}Emission properties of multiple atomic systems}
The same formalism can be applied to the case where multiple two-level atomic systems, indexed with $\alpha$, share the same photonic environment. The systems do not need to be identical, but we assume the separations of their transition frequencies $\omega_\alpha$ to be small with respect to the scale of spectral modulations of the properties of the photonic environment. This assumption will be relevant for the Markovian approximation. 
We additionally assume that the systems do not directly interact. However, the shared environment can be a carrier of interatomic coupling, as we will see below.

In the case of multiple atomic systems, the Hamiltonian from Eq.~(\ref{eq:hamiltonian}) should be generalized to the form
\begin{equation}\label{eq:hamiltonian_multi}
\mathcal{H} = \mathcal{H}_{\mathrm{field}} + \sum_\alpha\mathcal{H}_{\alpha} + \sum_\alpha \mathcal{H}_{\mathrm{int},\alpha},
\end{equation}
where $\mathcal{H}_{\alpha}=\hbar\omega_\alpha\sigma_\alpha^\dagger\sigma_\alpha$, and $\mathcal{H}_{\mathrm{int},\alpha}$ is given by Eq.~(\ref{eq:hamiltonian}) with the operator $\sigma$ replaced with $\sigma_\alpha$ and fields evaluated at positions $\mathbf{r}_\alpha$ of the $\alpha\textsuperscript{th}$ atomic system $\mathbf{E}^{(\pm)}(\mathbf{r}_0)\rightarrow \mathbf{E}^{(\pm)}_\alpha\equiv \mathbf{E}^{(\pm)}(\mathbf{r}_\alpha)$, $\mathbf{B}^{(\pm)}(\mathbf{r}_0)\rightarrow \mathbf{B}^{(\pm)}_\alpha\equiv\mathbf{B}^{(\pm)}(\mathbf{r}_\alpha)$. 
A set of Pauli operators $\sigma_\alpha$, $\sigma_{z,\alpha}$ describes the $\alpha\textsuperscript{th}$ atomic system. Different systems are naturally independent of each other, so the commutation rules for Pauli operators read $\left[\sigma_\alpha,\sigma_\beta\right]=0$,
$\left[\sigma_\alpha,\sigma_\beta^\dagger\right]=-\sigma_{z,\alpha}\delta_{\alpha\beta}$,
$\left[\sigma_\alpha,\sigma_{z,\beta}\right]=2\sigma_\alpha\delta_{\alpha\beta}$.

Steps to derive evolution equations of atomic operators in the Markovian approximation are listed in Appendix \ref{app:multiple_systems}. The resulting equations read
\begin{widetext}
\begin{eqnarray}\label{eq:sigma_multiatom}
\dot{\sigma}_{\beta}&=&-\left[i\left(\bar{\omega}+\delta_{\beta}\right)+\gamma_{\beta\beta}\right]\sigma_{\beta}+\sum_{\alpha\neq \beta}(i\xi_{\alpha\beta}+\gamma_{\alpha\beta})\sigma_{z,\beta}\sigma_\alpha
-\frac{i}{\hbar}\sigma_{z,\beta}\left[\mathbf{Q}^{\dagger\beta}:\nabla\mathbf{E}_{\beta,\mathrm{free}}^{(+)}+\mathbf{m}_{\beta}^{\dagger}\cdot\mathbf{B}_{\beta,\mathrm{free}}^{(+)}+\mathbf{d}_{\beta}^{\dagger}\cdot\mathbf{E}_{\beta,\mathrm{free}}^{(+)}\right]
\\
\dot{\sigma}_{z,\beta}&=&-\gamma_{\beta\beta}\left(\sigma_{z,\beta}+\mathds{1}\right)
+2i\left(\xi_{\alpha\beta}\sigma_\alpha^\dagger\sigma_\beta-\xi_{\alpha\beta}^\star\sigma_\beta^\dagger\sigma_\alpha \right)
\label{eq:sigmaz_multiatom}\\
&&+\frac{2i}{\hbar} \sigma_{\beta}^{\dagger}\left[\mathbf{d}_{\beta}^{\dagger}\cdot\mathbf{E}^{(+)}_{\beta,\mathrm{free}}+\mathbf{m}_{\beta}^{\dagger}\cdot\mathbf{B}^{(+)}_{\beta,\mathrm{free}}+\mathbf{Q}_{\beta}^{\dagger}:\nabla\mathbf{E}_{\beta,\mathrm{free}}^{(+)}\right]
-\frac{2i}{\hbar}
\left[\mathbf{E}^{(-)}_{\beta,\mathrm{free}}\cdot\mathbf{d}_{\beta}+\mathbf{B}^{(-)}_{\beta,\mathrm{free}}\cdot\mathbf{m}_{\beta}+\nabla\mathbf{E}^{(-)}_{\beta,\mathrm{free}}:\boldsymbol{Q}_{\beta}\right]\sigma_{\beta}.\nonumber
\end{eqnarray}
\end{widetext}

For a better understanding, it is useful to note that the same equations can be derived for a collection of atomic systems described by an effective Hamiltonian of the form 
\begin{equation}\label{eq:hamiltonian_eff}
\mathcal{H}_{\mathrm{eff}}=\sum_{\alpha}\hbar\left(\bar{\omega}+\delta_{\alpha}\right)\sigma_{\alpha}^{\dagger}\sigma_{\alpha}+\hbar\sum_{\alpha>\beta}\left(\xi_{\alpha\beta}\sigma_{\alpha}^{\dagger}\sigma_{\beta}+\xi_{\alpha\beta}^{\star}\sigma_{\beta}^{\dagger}\sigma_{\alpha}\right).
\end{equation}
In the Hamiltonian in Eq.~(\ref{eq:hamiltonian_multi}), the photonic environment explicitly plays the role of the interaction carrier. In (\ref{eq:hamiltonian_eff}) the environment is eliminated, and an effective and direct multipole-multipole coupling is present instead with a strength 

\begin{eqnarray}\label{eq:hamiltonian_effective}
\xi_{\alpha\beta}&=&\sum_{mn}\left\{ \mathcal{P}\int_{0}^{\infty}d\omega R_{mn}\left(\omega\right)\frac{\omega^{2}}{\omega-\bar{\omega}} \mathrm{Im}\,G_{mn}\left(\mathbf{r}^{\prime},\mathbf{r},\omega\right)|_{\substack{\mathbf{r}=\mathbf{r_{\beta}},\\
\mathbf{r}^{\prime}=\mathbf{r}_{\alpha}
}}\right.\nonumber\\
&&\left.+I_{mn}\left(\bar{\omega}\right)\pi\bar{\omega}^{2} \mathrm{Im}\,G_{mn}\left(\mathbf{r}^{\prime},\mathbf{r},\bar{\omega}\right)
|_{\substack{\mathbf{r}=\mathbf{r_{\beta}},\\ \mathbf{r}^{\prime}=\mathbf{r}_{\alpha}}}
\right\}, 
\end{eqnarray}
where $R_{mn}(\omega)$ and $I_{mn}(\omega)$ are expressed through multipolar elements and differential operators, as given in Appendix \ref{app:multiple_systems}. 
From Eq.~(\ref{eqS:PV_replacement}) of Appendix \ref{app:kramers} it follows that if the transition frequency is sufficiently large, the expression for the coupling can be simplified to the form
\begin{eqnarray}
\xi_{\alpha\beta}&=&\pi\bar{\omega}^2\sum_{mn}\left\{  R_{mn}(\bar{\omega}) \mathrm{Re}\,G_{mn}\left(\mathbf{r}^{\prime},\mathbf{r},\bar{\omega}\right)
|_{\substack{\mathbf{r}=\mathbf{r_{\beta}},\\ \mathbf{r}^{\prime}=\mathbf{r}_{\alpha}}}\right.\nonumber\\
&&+\left.I_{mn}\left(\bar{\omega}\right) \mathrm{Im}\,G_{mn}\left(\mathbf{r}^{\prime},\mathbf{r},\bar{\omega}\right)
|_{\substack{\mathbf{r}=\mathbf{r_{\beta}},\\ \mathbf{r}^{\prime}=\mathbf{r}_{\alpha}}}
\right\} \\
&&+\int_0^\infty \frac{\omega^2\bar{\omega}}{\omega^2+\bar{\omega}^2} R_{mn}(i\omega)\mathrm{Re}\,G_{mn}\left(\mathbf{r}^{\prime},\mathbf{r},i\omega\right)
|_{\substack{\mathbf{r}=\mathbf{r_{\beta}},\\ \mathbf{r}^{\prime}=\mathbf{r}_{\alpha}}},\nonumber
\end{eqnarray}
where the principal value integral is no longer present and the integration is now performed along the imaginary axis. There, the Green's tensor shows greater numerical stability due to its decaying rather than oscillating character.

The multipole-multipole interaction strength $\xi_{\alpha\beta}$ is a generalization of the dipole-dipole coupling which in free space scales as $R^{-3}$, $R$ being the interatomic distance. Contrary, a modified photonic environment, e.g., in a photonic crystal, near a nanoparticle or a nanowire, might allow not only for stronger interactions but also for extended interaction distances \cite{Dzsotjan2010}. Due to the enhancement of off-diagonal elements of the Green's tensor, long-range coupling of multipoles or arbitrary, in general non parallel orientations, may be enabled.
Due to the strong field localization, corrections beyond the electric dipole approximation in such systems may be significant, and coupling of different multipoles is possible. Interference of different interaction components may lead to further enhancement or suppression of interaction strength, resulting in a corresponding modification of interaction distance.
Please note that due to the large width of the peak in the density of states, assumed in Appendix \ref{app:multiple_systems}, atomic systems with slightly different transition frequencies may in general be coupled. 

Dissipators in Eqs.~(\ref{eq:sigma_multiatom}) include emission rates of an individual, $\alpha\textsuperscript{th}$ atomic system $\gamma_{\alpha\alpha}$ (reducing to $\gamma$  from the previous Section in case of a single system), and collective decay rates $\gamma_{\alpha\beta}$. They arise because each atomic system is capable of modifying the photonic environment of the others and they are defined as
\begin{eqnarray}\label{eq:gamma_collective}
\gamma_{\alpha\beta} &=& 
2\sum_{mj}\left\{ R_{mj}\left(\bar{\omega}\right)\pi\bar{\omega}^{2}\mathrm{Im}\,G_{mj}\left(\mathbf{r}^{\prime},\mathbf{r},\bar{\omega}\right)
|_{\substack{\mathbf{r}=\mathbf{r_{\beta}},\\ \mathbf{r}^{\prime}=\mathbf{r}_{\alpha}}}\right.\\
&&-\left.\mathcal{P}\int_{0}^{\infty}d\omega I_{mj}\left(\omega\right)\frac{\omega^{2}}{\omega-\bar{\omega}}
\mathrm{Im}\,G_{mj}\left(\mathbf{r}^{\prime},\mathbf{r},\omega\right)
|_{\substack{\mathbf{r}=\mathbf{r_{\beta}},\\ \mathbf{r}^{\prime}=\mathbf{r}_{\alpha}}}
\right\},\nonumber
\end{eqnarray}
valid also for $\beta=\alpha$, which yields $I_{mj}\left(\omega\right)=0$. Again, using Eq.~(\ref{eqS:PV_replacement}) of Appendix \ref{app:kramers} we can simplify the expression to
\begin{eqnarray}
\gamma_{\alpha\beta} &=& 
2\pi\bar{\omega}^{2}\sum_{mn}\left\{ R_{mn}\left(\bar{\omega}\right)\mathrm{Im}\,G_{mn}\left(\mathbf{r}^{\prime},\mathbf{r},\bar{\omega}\right)
|_{\substack{\mathbf{r}=\mathbf{r_{\beta}},\\ \mathbf{r}^{\prime}=\mathbf{r}_{\alpha}}}\right.\nonumber\\
&&-\left.I_{mn}\left(\bar{\omega} \right)\mathrm{Re}\,G_{mn}\left(\mathbf{r}^{\prime},\mathbf{r},\bar{\omega}\right)
|_{\substack{\mathbf{r}=\mathbf{r_{\beta}},\\ \mathbf{r}^{\prime}=\mathbf{r}_{\alpha}}}
\right\}\\
&&-2\int_0^\infty d\omega \frac{\omega^2\bar{\omega}}{\omega^2+\bar{\omega}^2}\mathrm{Re}\left[I_{mn}\left(i\omega\right)G_{mn}\left(\mathbf{r}^{\prime},\mathbf{r},\omega\right)
|_{\substack{\mathbf{r}=\mathbf{r_{\beta}},\\ \mathbf{r}^{\prime}=\mathbf{r}_{\alpha}}}\right]
.\nonumber
\end{eqnarray}
Please note that for identical emitters this problem can be discussed in terms of Dicke superradiance, and would be a straightforward generalization of results of Ref.\cite{Sinha2018}.

\section{Examples} \label{sec:example}
In this section we evaluate the formulas derived above first to the case of a homogeneous background dielectric and second to an exemplary selected nanostructured environment into which the atomic system is placed. In the first considered case, our goal is to retrieve familiar scaling of different multipolar components of spontaneous emission rates with different powers of refractive index \cite{Lukosz1977, Henderson2006}. In the latter case, we demonstrate that in a suitably engineered environment contributions beyond electric dipole may have a significant impact on atomic system's dynamics. 
\subsection{Homogeneous background medium}
In a homogeneous, isotropic, and infinitely extended medium the Green's tensor takes the form
\begin{equation}\label{eq:green_homogeneous}
\textbf{G}(\mathbf{R},\omega) = \left(\mathbf{1} + \frac{ikR-1}{k^2R^2} \textbf{1} + \frac{3-3ikR-k^2R^2}{k^2R^4}\mathbf{RR}\right)\frac{e^{ikR}}{4\pi R},
\end{equation}
where $R=\left| \mathbf{R}\right|=\left| \mathbf{r}-\mathbf{r}^\prime\right|$ is the distance between the source and observation point where the field is to be evaluated, and the wave number in the homogenous medium reads $k=\frac{\omega}{c}\sqrt{\epsilon(\omega)}=\frac{\omega}{c}n(\omega)$, $n(\omega)$ being a position-independent refractive index. 

In Appendix \ref{app:homogeneous} we show that away from the medium resonances the imaginary part of the Green's tensor is diagonal in the  limit of $R\rightarrow 0$ with
\begin{equation}
    \lim_{R\rightarrow 0}\mathrm{Im}\, G_{jk}(\mathbf{R},\omega) = \frac{k^3}{60\pi}R_jR_k,
\end{equation}
for $k\neq j$, and is exactly $0$ for $R=0$. The diagonal elements however, are finite and read as
\begin{equation}
    \mathrm{Im}\,G_{jj}(\mathbf{R},\omega) = \frac{k}{6\pi}-\frac{k^3}{48\pi}R^2+\frac{k^3}{60\pi}R_jR_j+O(R^4).
\end{equation}
Inserting the limit at $R\rightarrow 0$ in Eq.~\ref{eq:gamma_main}, we retrieve the Weisskopf-Wigner result for the electric dipole contribution to spontaneous emission
\begin{equation}
    \gamma_{\mathrm{ED}} = \frac{n \omega_0^3 |d|^2}{3 \pi \hbar\epsilon_0 c^3}.
\end{equation}
Higher-order terms may be evaluated based on derivatives found in Eqs.~(\ref{eqS:2nd_derivative_offdiagonal} \& \ref{eqS:2nd_derivative_diagonal}) of Appendix \ref{app:homogeneous}
\begin{eqnarray}
    \gamma_{\mathrm{MD}} &=& \frac{n^3\omega^3\left|\mathbf{m}\right|^2}{3\pi\hbar\epsilon_0 c^5}, \label{eq:gamma_md_fs}\\
    \gamma_{\mathrm{EQ}} &=& \frac{n^3\omega^5\sum_{mn}\left|Q_{mn}\right|^2}{10\pi\hbar\epsilon_0 c^5}. \label{eq:gamma_eq_fs}
\end{eqnarray}
As the result, we find the familiar result for transition rates \textit{via} higher-order channels \cite{Lukosz1977,Craig1998}, which scale with the third power of the refractive index \cite{Lukosz1977,Henderson2006}.
Similarly, multipole-multipole interaction terms can be retrieved. 

\subsection{Pair of metallic nanospheres}
\begin{figure}
            \includegraphics[width=\linewidth]{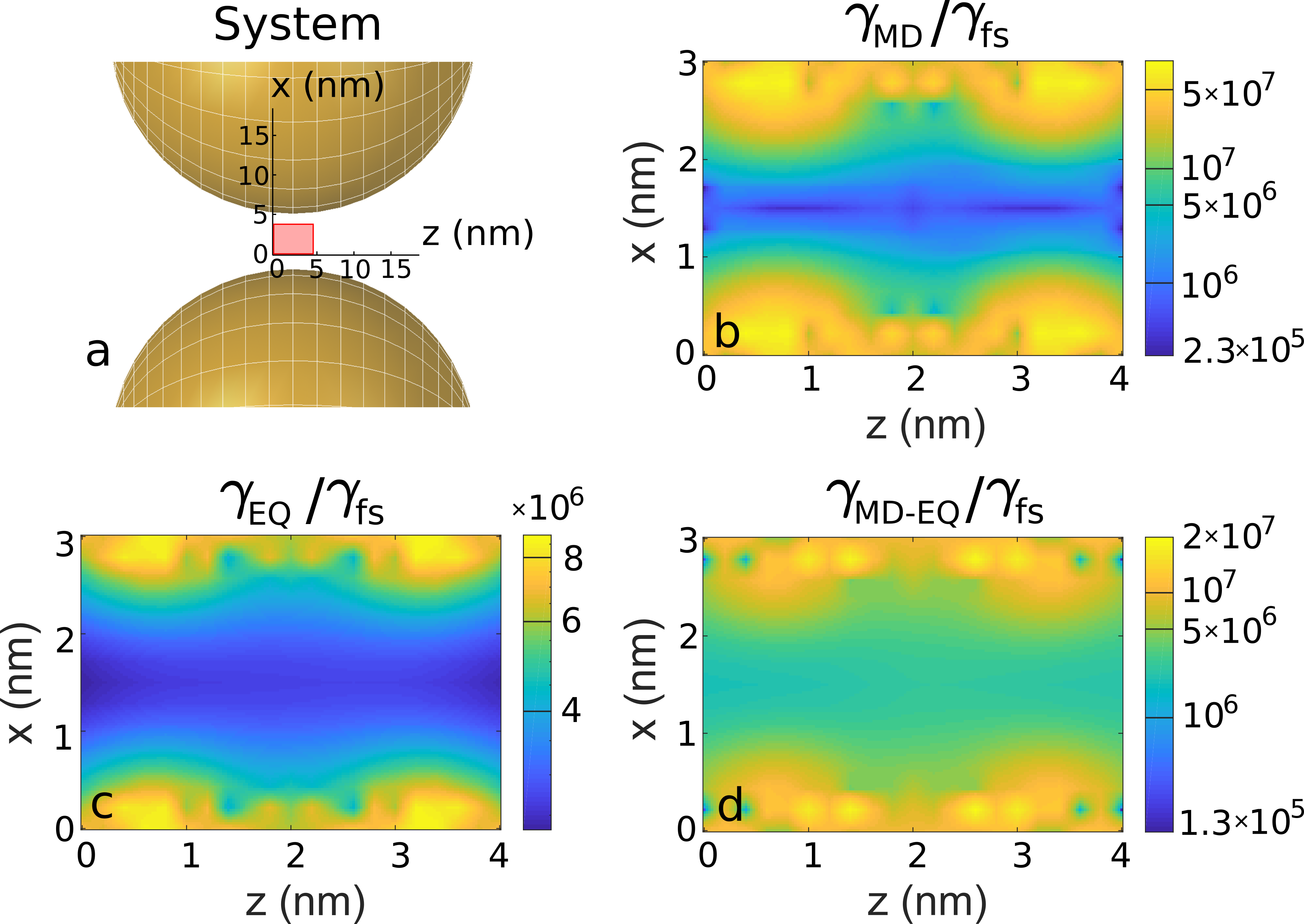}
        \caption[aa]{\small System scheme and emission enhancement due to various multipole channels. (a) Exemplary system for calculating emission enhancement - two silver nanospheres of 40 nm diameter, separated by a 6 nm gap. The pink rectangle indicates the grid on which Green's tensor is evaluated. (b-d) Spontaneous emission enhancement from the MD-MD channel (b), EQ-EQ channel (c) and interference between MD and EQ channels (d) normalized to the free space value. \label{fig:system_scheme}} 
    \end{figure}

To offer also an example of a structured photonic environment, we consider here a pair of silver nanospheres of 40 nm diameter, separated by a 6 nm gap inside of which an atomic system is positioned.
The chosen coordinate frame is shown in Fig.~\ref{fig:system_scheme}(a). 
The Green's tensor was calculated using the MNPBEM toolbox for MATLAB \cite{mnpbem}: a Maxwell equation solver based on the Boundary Element Method \cite{deAbajo2002,Hohenester2008,deAbajo2010}. Full Maxwell equations were solved. 
Silver was modeled using data from Ref.~\onlinecite{johnson}. 
The tensor is calculated at the frequency $\omega_0=2\pi\times 789$ $\mathrm{ps}^{-1}$ on a $4\,\mathrm{nm}\times 3\,\mathrm{nm}$ grid located in the symmetry plane $y=0$ between the nanospheres, as marked by the rectangle in Fig.~\ref{fig:system_scheme}(a). 
The Green's tensor's elements and derivatives are presented in the \textit{Supplementary Material}.

We now consider a two-level atomic system with transition frequency $\omega_0$. For simplicity we choose the transition electric dipole moment to be vanishing. The magnetic transition dipole moment parallel to the $z$ axis $m_z = 2i\mu_B$, $\mu_B$ standing for Bohr magneton, and the electric transition quadrupole moment in the $xy$ plane $Q_{xy} = Q_{yx} = ea_0^2$ with elementary charge $e$ and Bohr radius $a_0$. The chosen values correspond to both moments equal to 1 atomic unit, i.e. values characteristic for atoms and molecules. The transition rates depend on the position of the atomic system with respect to the nanospheres, and are shown in Figs.~\ref{fig:system_scheme}(b-d). They have been normalized to the free-space value $\gamma_\mathrm{fs}=\gamma_\mathrm{MD,fs}+\gamma_\mathrm{EQ,fs}$, where $\gamma_\mathrm{MD,fs}$ and $\gamma_\mathrm{EQ,fs}$ are respectively given by Eqs.~(\ref{eq:gamma_md_fs}) and (\ref{eq:gamma_eq_fs}) with $n=1$. We only consider atomic system's positions in the rectangle from Fig.~\ref{fig:system_scheme}(a). In general, the enhanced transition rates in the higher-order channels exceed the free-space value $\gamma_\mathrm{fs}$ by at least $6$ orders of magnitude. Please note that the resulting rate is enhanced to the MHz level and becomes comparable to typical free-space values of atomic transition rates of the electric dipole channel with a typical value of dipole moment $d=ea_0$. Among possible applications this suggests potential for enhancement of optical activity, in particular circular dichroism. 
Among the two considered higher-order channels, the magnetic dipole transition channel dominates by two orders of magnitude over the electric quadrupole one. However, the latter is manifested through interference, which we find always destructive in the investigated region, and whose contribution to the total transition rate is of the order of $10\%$.

\section{Discussion}
We have studied dynamics of atomic systems coupled to a photonic environment in its vacuum state. The environment is described in terms of the electromagnetic Green’s tensor, and in the interaction contributions beyond the paradigmatic electric dipole approximation have been included. The derived formalism allows to evaluate dynamical parameters characterizing optical properties of atomic systems: both the individual and the collective contributions to energy shifts and decay rates. Inclusion of terms beyond the electric dipole approximation allows to study role of higher multipolar channels, including enhancement or suppression through interference of different interaction mechanisms. Examples of phenomena to be investigated are optical activity, multipole-multipole interactions between atomic systems and spontaneous emission suppression due to interference of different mutipolar channels in tailored photonic environments.

\section*{Acknowledgements}
MK \& KS acknowledge support from the Foundation for Polish Science (Project HEIMaT No. Homing/2016-1/8) within the European Regional Development Fund. CR and IF acknowledge support by the Deutsche Forschungsgemeinschaft (DFG, German Research Foundation) - project number 378579271 - within project RO 3640/8-1.
OB is grateful for the support of the Toru\'{n} Astrophysics/Physics Summer Program TAPS 2018. 

\bibliography{biblio}

\begin{thebibliography}{49}%
\makeatletter
\providecommand \@ifxundefined [1]{%
 \@ifx{#1\undefined}
}%
\providecommand \@ifnum [1]{%
 \ifnum #1\expandafter \@firstoftwo
 \else \expandafter \@secondoftwo
 \fi
}%
\providecommand \@ifx [1]{%
 \ifx #1\expandafter \@firstoftwo
 \else \expandafter \@secondoftwo
 \fi
}%
\providecommand \natexlab [1]{#1}%
\providecommand \enquote  [1]{``#1''}%
\providecommand \bibnamefont  [1]{#1}%
\providecommand \bibfnamefont [1]{#1}%
\providecommand \citenamefont [1]{#1}%
\providecommand \href@noop [0]{\@secondoftwo}%
\providecommand \href [0]{\begingroup \@sanitize@url \@href}%
\providecommand \@href[1]{\@@startlink{#1}\@@href}%
\providecommand \@@href[1]{\endgroup#1\@@endlink}%
\providecommand \@sanitize@url [0]{\catcode `\\12\catcode `\$12\catcode
  `\&12\catcode `\#12\catcode `\^12\catcode `\_12\catcode `\%12\relax}%
\providecommand \@@startlink[1]{}%
\providecommand \@@endlink[0]{}%
\providecommand \url  [0]{\begingroup\@sanitize@url \@url }%
\providecommand \@url [1]{\endgroup\@href {#1}{\urlprefix }}%
\providecommand \urlprefix  [0]{URL }%
\providecommand \Eprint [0]{\href }%
\providecommand \doibase [0]{http://dx.doi.org/}%
\providecommand \selectlanguage [0]{\@gobble}%
\providecommand \bibinfo  [0]{\@secondoftwo}%
\providecommand \bibfield  [0]{\@secondoftwo}%
\providecommand \translation [1]{[#1]}%
\providecommand \BibitemOpen [0]{}%
\providecommand \bibitemStop [0]{}%
\providecommand \bibitemNoStop [0]{.\EOS\space}%
\providecommand \EOS [0]{\spacefactor3000\relax}%
\providecommand \BibitemShut  [1]{\csname bibitem#1\endcsname}%
\let\auto@bib@innerbib\@empty
\bibitem [{\citenamefont {Purcell}(1946)}]{purcell1946}%
  \BibitemOpen
  \bibfield  {author} {\bibinfo {author} {\bibfnamefont {E.}~\bibnamefont
  {Purcell}},\ }\href@noop {} {\bibfield  {journal} {\bibinfo  {journal} {Phys.
  Rev.}\ }\textbf {\bibinfo {volume} {69}},\ \bibinfo {pages} {37} (\bibinfo
  {year} {1946})}\BibitemShut {NoStop}%
\bibitem [{\citenamefont {G{\'e}rard}\ \emph {et~al.}(1998)\citenamefont
  {G{\'e}rard}, \citenamefont {Sermage}, \citenamefont {Gayral}, \citenamefont
  {Legrand}, \citenamefont {Costard},\ and\ \citenamefont
  {Thierry-Mieg}}]{Gerard1998}%
  \BibitemOpen
  \bibfield  {author} {\bibinfo {author} {\bibfnamefont {J.}~\bibnamefont
  {G{\'e}rard}}, \bibinfo {author} {\bibfnamefont {B.}~\bibnamefont {Sermage}},
  \bibinfo {author} {\bibfnamefont {B.}~\bibnamefont {Gayral}}, \bibinfo
  {author} {\bibfnamefont {B.}~\bibnamefont {Legrand}}, \bibinfo {author}
  {\bibfnamefont {E.}~\bibnamefont {Costard}}, \ and\ \bibinfo {author}
  {\bibfnamefont {V.}~\bibnamefont {Thierry-Mieg}},\ }\href@noop {} {\bibfield
  {journal} {\bibinfo  {journal} {Physical Review Letters}\ }\textbf {\bibinfo
  {volume} {81}},\ \bibinfo {pages} {1110} (\bibinfo {year}
  {1998})}\BibitemShut {NoStop}%
\bibitem [{\citenamefont {Englund}\ \emph {et~al.}(2005)\citenamefont
  {Englund}, \citenamefont {Fattal}, \citenamefont {Waks}, \citenamefont
  {Solomon}, \citenamefont {Zhang}, \citenamefont {Nakaoka}, \citenamefont
  {Arakawa}, \citenamefont {Yamamoto},\ and\ \citenamefont
  {Vu{\v{c}}kovi{\'c}}}]{Englund2005}%
  \BibitemOpen
  \bibfield  {author} {\bibinfo {author} {\bibfnamefont {D.}~\bibnamefont
  {Englund}}, \bibinfo {author} {\bibfnamefont {D.}~\bibnamefont {Fattal}},
  \bibinfo {author} {\bibfnamefont {E.}~\bibnamefont {Waks}}, \bibinfo {author}
  {\bibfnamefont {G.}~\bibnamefont {Solomon}}, \bibinfo {author} {\bibfnamefont
  {B.}~\bibnamefont {Zhang}}, \bibinfo {author} {\bibfnamefont
  {T.}~\bibnamefont {Nakaoka}}, \bibinfo {author} {\bibfnamefont
  {Y.}~\bibnamefont {Arakawa}}, \bibinfo {author} {\bibfnamefont
  {Y.}~\bibnamefont {Yamamoto}}, \ and\ \bibinfo {author} {\bibfnamefont
  {J.}~\bibnamefont {Vu{\v{c}}kovi{\'c}}},\ }\href@noop {} {\bibfield
  {journal} {\bibinfo  {journal} {Physical Review Letters}\ }\textbf {\bibinfo
  {volume} {95}},\ \bibinfo {pages} {013904} (\bibinfo {year}
  {2005})}\BibitemShut {NoStop}%
\bibitem [{\citenamefont {Yuan}\ \emph {et~al.}(2013)\citenamefont {Yuan},
  \citenamefont {Khatua}, \citenamefont {Zijlstra}, \citenamefont {Yorulmaz},\
  and\ \citenamefont {Orrit}}]{Yuan2013}%
  \BibitemOpen
  \bibfield  {author} {\bibinfo {author} {\bibfnamefont {H.}~\bibnamefont
  {Yuan}}, \bibinfo {author} {\bibfnamefont {S.}~\bibnamefont {Khatua}},
  \bibinfo {author} {\bibfnamefont {P.}~\bibnamefont {Zijlstra}}, \bibinfo
  {author} {\bibfnamefont {M.}~\bibnamefont {Yorulmaz}}, \ and\ \bibinfo
  {author} {\bibfnamefont {M.}~\bibnamefont {Orrit}},\ }\href@noop {}
  {\bibfield  {journal} {\bibinfo  {journal} {Angewandte Chemie International
  Edition}\ }\textbf {\bibinfo {volume} {52}},\ \bibinfo {pages} {1217}
  (\bibinfo {year} {2013})}\BibitemShut {NoStop}%
\bibitem [{\citenamefont {Akselrod}\ \emph {et~al.}(2014)\citenamefont
  {Akselrod}, \citenamefont {Argyropoulos}, \citenamefont {Hoang},
  \citenamefont {Cirac{\`\i}}, \citenamefont {Fang}, \citenamefont {Huang},
  \citenamefont {Smith},\ and\ \citenamefont {Mikkelsen}}]{Akselrod2014}%
  \BibitemOpen
  \bibfield  {author} {\bibinfo {author} {\bibfnamefont {G.~M.}\ \bibnamefont
  {Akselrod}}, \bibinfo {author} {\bibfnamefont {C.}~\bibnamefont
  {Argyropoulos}}, \bibinfo {author} {\bibfnamefont {T.~B.}\ \bibnamefont
  {Hoang}}, \bibinfo {author} {\bibfnamefont {C.}~\bibnamefont {Cirac{\`\i}}},
  \bibinfo {author} {\bibfnamefont {C.}~\bibnamefont {Fang}}, \bibinfo {author}
  {\bibfnamefont {J.}~\bibnamefont {Huang}}, \bibinfo {author} {\bibfnamefont
  {D.~R.}\ \bibnamefont {Smith}}, \ and\ \bibinfo {author} {\bibfnamefont
  {M.~H.}\ \bibnamefont {Mikkelsen}},\ }\href@noop {} {\bibfield  {journal}
  {\bibinfo  {journal} {Nature Photonics}\ }\textbf {\bibinfo {volume} {8}},\
  \bibinfo {pages} {835} (\bibinfo {year} {2014})}\BibitemShut {NoStop}%
\bibitem [{\citenamefont {Ficek}\ \emph {et~al.}(1987)\citenamefont {Ficek},
  \citenamefont {Tana{\'s}},\ and\ \citenamefont {Kielich}}]{Ficek1987}%
  \BibitemOpen
  \bibfield  {author} {\bibinfo {author} {\bibfnamefont {Z.}~\bibnamefont
  {Ficek}}, \bibinfo {author} {\bibfnamefont {R.}~\bibnamefont {Tana{\'s}}}, \
  and\ \bibinfo {author} {\bibfnamefont {S.}~\bibnamefont {Kielich}},\
  }\href@noop {} {\bibfield  {journal} {\bibinfo  {journal} {Physica A:
  Statistical Mechanics and its Applications}\ }\textbf {\bibinfo {volume}
  {146}},\ \bibinfo {pages} {452} (\bibinfo {year} {1987})}\BibitemShut
  {NoStop}%
\bibitem [{\citenamefont {Dicke}(1954)}]{Dicke1954}%
  \BibitemOpen
  \bibfield  {author} {\bibinfo {author} {\bibfnamefont {R.~H.}\ \bibnamefont
  {Dicke}},\ }\href@noop {} {\bibfield  {journal} {\bibinfo  {journal}
  {Physical Review}\ }\textbf {\bibinfo {volume} {93}},\ \bibinfo {pages} {99}
  (\bibinfo {year} {1954})}\BibitemShut {NoStop}%
\bibitem [{\citenamefont {Gross}\ and\ \citenamefont
  {Haroche}(1982)}]{Gross1982}%
  \BibitemOpen
  \bibfield  {author} {\bibinfo {author} {\bibfnamefont {M.}~\bibnamefont
  {Gross}}\ and\ \bibinfo {author} {\bibfnamefont {S.}~\bibnamefont
  {Haroche}},\ }\href@noop {} {\bibfield  {journal} {\bibinfo  {journal}
  {Physics Reports}\ }\textbf {\bibinfo {volume} {93}},\ \bibinfo {pages} {301}
  (\bibinfo {year} {1982})}\BibitemShut {NoStop}%
\bibitem [{\citenamefont {Dzsotjan}\ \emph {et~al.}(2010)\citenamefont
  {Dzsotjan}, \citenamefont {S\o{}rensen},\ and\ \citenamefont
  {Fleischhauer}}]{Dzsotjan2010}%
  \BibitemOpen
  \bibfield  {author} {\bibinfo {author} {\bibfnamefont {D.}~\bibnamefont
  {Dzsotjan}}, \bibinfo {author} {\bibfnamefont {A.~S.}\ \bibnamefont
  {S\o{}rensen}}, \ and\ \bibinfo {author} {\bibfnamefont {M.}~\bibnamefont
  {Fleischhauer}},\ }\href
  {https://link.aps.org/doi/10.1103/PhysRevB.82.075427} {\bibfield  {journal}
  {\bibinfo  {journal} {Phys. Rev. B}\ }\textbf {\bibinfo {volume} {82}},\
  \bibinfo {pages} {075427} (\bibinfo {year} {2010})}\BibitemShut {NoStop}%
\bibitem [{\citenamefont {Bouchet}\ and\ \citenamefont
  {Carminati}(2019)}]{Bouchet2019}%
  \BibitemOpen
  \bibfield  {author} {\bibinfo {author} {\bibfnamefont {D.}~\bibnamefont
  {Bouchet}}\ and\ \bibinfo {author} {\bibfnamefont {R.}~\bibnamefont
  {Carminati}},\ }\href@noop {} {\bibfield  {journal} {\bibinfo  {journal} {J.
  Opt. Soc. Am. A}\ }\textbf {\bibinfo {volume} {36}},\ \bibinfo {pages} {186}
  (\bibinfo {year} {2019})}\BibitemShut {NoStop}%
\bibitem [{\citenamefont {Konzelmann}\ \emph {et~al.}(2019)\citenamefont
  {Konzelmann}, \citenamefont {Kr{\"u}ger},\ and\ \citenamefont
  {Giessen}}]{Konzelmann2019}%
  \BibitemOpen
  \bibfield  {author} {\bibinfo {author} {\bibfnamefont {A.~M.}\ \bibnamefont
  {Konzelmann}}, \bibinfo {author} {\bibfnamefont {S.~O.}\ \bibnamefont
  {Kr{\"u}ger}}, \ and\ \bibinfo {author} {\bibfnamefont {H.}~\bibnamefont
  {Giessen}},\ }\href@noop {} {\bibfield  {journal} {\bibinfo  {journal} {arXiv
  preprint arXiv:1905.07131}\ } (\bibinfo {year} {2019})}\BibitemShut {NoStop}%
\bibitem [{\citenamefont {Schuller}\ \emph {et~al.}(2010)\citenamefont
  {Schuller}, \citenamefont {Barnard}, \citenamefont {Cai}, \citenamefont
  {Jun}, \citenamefont {White},\ and\ \citenamefont
  {Brongersma}}]{Schuller2010}%
  \BibitemOpen
  \bibfield  {author} {\bibinfo {author} {\bibfnamefont {J.~A.}\ \bibnamefont
  {Schuller}}, \bibinfo {author} {\bibfnamefont {E.~S.}\ \bibnamefont
  {Barnard}}, \bibinfo {author} {\bibfnamefont {W.}~\bibnamefont {Cai}},
  \bibinfo {author} {\bibfnamefont {Y.~C.}\ \bibnamefont {Jun}}, \bibinfo
  {author} {\bibfnamefont {J.~S.}\ \bibnamefont {White}}, \ and\ \bibinfo
  {author} {\bibfnamefont {M.~L.}\ \bibnamefont {Brongersma}},\ }\href@noop {}
  {\bibfield  {journal} {\bibinfo  {journal} {Nature Materials}\ }\textbf
  {\bibinfo {volume} {9}},\ \bibinfo {pages} {193} (\bibinfo {year}
  {2010})}\BibitemShut {NoStop}%
\bibitem [{\citenamefont {Gramotnev}\ and\ \citenamefont
  {Bozhevolnyi}(2010)}]{Gramotnev2010}%
  \BibitemOpen
  \bibfield  {author} {\bibinfo {author} {\bibfnamefont {D.~K.}\ \bibnamefont
  {Gramotnev}}\ and\ \bibinfo {author} {\bibfnamefont {S.~I.}\ \bibnamefont
  {Bozhevolnyi}},\ }\href@noop {} {\bibfield  {journal} {\bibinfo  {journal}
  {Nature Photonics}\ }\textbf {\bibinfo {volume} {4}},\ \bibinfo {pages} {83}
  (\bibinfo {year} {2010})}\BibitemShut {NoStop}%
\bibitem [{\citenamefont {Benz}\ \emph {et~al.}(2016)\citenamefont {Benz},
  \citenamefont {Schmidt}, \citenamefont {Dreismann}, \citenamefont
  {Chikkaraddy}, \citenamefont {Zhang}, \citenamefont {Demetriadou},
  \citenamefont {Carnegie}, \citenamefont {Ohadi}, \citenamefont {de~Nijs},
  \citenamefont {Esteban} \emph {et~al.}}]{Benz2016}%
  \BibitemOpen
  \bibfield  {author} {\bibinfo {author} {\bibfnamefont {F.}~\bibnamefont
  {Benz}}, \bibinfo {author} {\bibfnamefont {M.~K.}\ \bibnamefont {Schmidt}},
  \bibinfo {author} {\bibfnamefont {A.}~\bibnamefont {Dreismann}}, \bibinfo
  {author} {\bibfnamefont {R.}~\bibnamefont {Chikkaraddy}}, \bibinfo {author}
  {\bibfnamefont {Y.}~\bibnamefont {Zhang}}, \bibinfo {author} {\bibfnamefont
  {A.}~\bibnamefont {Demetriadou}}, \bibinfo {author} {\bibfnamefont
  {C.}~\bibnamefont {Carnegie}}, \bibinfo {author} {\bibfnamefont
  {H.}~\bibnamefont {Ohadi}}, \bibinfo {author} {\bibfnamefont
  {B.}~\bibnamefont {de~Nijs}}, \bibinfo {author} {\bibfnamefont
  {R.}~\bibnamefont {Esteban}},  \emph {et~al.},\ }\href@noop {} {\bibfield
  {journal} {\bibinfo  {journal} {Science}\ }\textbf {\bibinfo {volume}
  {354}},\ \bibinfo {pages} {726} (\bibinfo {year} {2016})}\BibitemShut
  {NoStop}%
\bibitem [{\citenamefont {Schmidt}\ \emph {et~al.}(2012)\citenamefont
  {Schmidt}, \citenamefont {Esteban}, \citenamefont {S{\'a}enz}, \citenamefont
  {Su{\'a}rez-Lacalle}, \citenamefont {Mackowski},\ and\ \citenamefont
  {Aizpurua}}]{Schmidt2012}%
  \BibitemOpen
  \bibfield  {author} {\bibinfo {author} {\bibfnamefont {M.~K.}\ \bibnamefont
  {Schmidt}}, \bibinfo {author} {\bibfnamefont {R.}~\bibnamefont {Esteban}},
  \bibinfo {author} {\bibfnamefont {J.}~\bibnamefont {S{\'a}enz}}, \bibinfo
  {author} {\bibfnamefont {I.}~\bibnamefont {Su{\'a}rez-Lacalle}}, \bibinfo
  {author} {\bibfnamefont {S.}~\bibnamefont {Mackowski}}, \ and\ \bibinfo
  {author} {\bibfnamefont {J.}~\bibnamefont {Aizpurua}},\ }\href@noop {}
  {\bibfield  {journal} {\bibinfo  {journal} {Optics Express}\ }\textbf
  {\bibinfo {volume} {20}},\ \bibinfo {pages} {13636} (\bibinfo {year}
  {2012})}\BibitemShut {NoStop}%
\bibitem [{\citenamefont {Hein}\ and\ \citenamefont
  {Giessen}(2013)}]{Hein2013}%
  \BibitemOpen
  \bibfield  {author} {\bibinfo {author} {\bibfnamefont {S.~M.}\ \bibnamefont
  {Hein}}\ and\ \bibinfo {author} {\bibfnamefont {H.}~\bibnamefont {Giessen}},\
  }\href@noop {} {\bibfield  {journal} {\bibinfo  {journal} {Physical Review
  Letters}\ }\textbf {\bibinfo {volume} {111}},\ \bibinfo {pages} {026803}
  (\bibinfo {year} {2013})}\BibitemShut {NoStop}%
\bibitem [{\citenamefont {Karaveli}\ and\ \citenamefont
  {Zia}(2010)}]{Karaveli2010}%
  \BibitemOpen
  \bibfield  {author} {\bibinfo {author} {\bibfnamefont {S.}~\bibnamefont
  {Karaveli}}\ and\ \bibinfo {author} {\bibfnamefont {R.}~\bibnamefont {Zia}},\
  }\href@noop {} {\bibfield  {journal} {\bibinfo  {journal} {Optics Letters}\
  }\textbf {\bibinfo {volume} {35}},\ \bibinfo {pages} {3318} (\bibinfo {year}
  {2010})}\BibitemShut {NoStop}%
\bibitem [{\citenamefont {Taminiau}\ \emph {et~al.}(2012)\citenamefont
  {Taminiau}, \citenamefont {Karaveli}, \citenamefont {Van~Hulst},\ and\
  \citenamefont {Zia}}]{Taminiau2012}%
  \BibitemOpen
  \bibfield  {author} {\bibinfo {author} {\bibfnamefont {T.~H.}\ \bibnamefont
  {Taminiau}}, \bibinfo {author} {\bibfnamefont {S.}~\bibnamefont {Karaveli}},
  \bibinfo {author} {\bibfnamefont {N.~F.}\ \bibnamefont {Van~Hulst}}, \ and\
  \bibinfo {author} {\bibfnamefont {R.}~\bibnamefont {Zia}},\ }\href@noop {}
  {\bibfield  {journal} {\bibinfo  {journal} {Nature Communications}\ }\textbf
  {\bibinfo {volume} {3}},\ \bibinfo {pages} {979} (\bibinfo {year}
  {2012})}\BibitemShut {NoStop}%
\bibitem [{\citenamefont {Kasperczyk}\ \emph {et~al.}(2015)\citenamefont
  {Kasperczyk}, \citenamefont {Person}, \citenamefont {Ananias}, \citenamefont
  {Carlos},\ and\ \citenamefont {Novotny}}]{Kasperczyk2015}%
  \BibitemOpen
  \bibfield  {author} {\bibinfo {author} {\bibfnamefont {M.}~\bibnamefont
  {Kasperczyk}}, \bibinfo {author} {\bibfnamefont {S.}~\bibnamefont {Person}},
  \bibinfo {author} {\bibfnamefont {D.}~\bibnamefont {Ananias}}, \bibinfo
  {author} {\bibfnamefont {L.~D.}\ \bibnamefont {Carlos}}, \ and\ \bibinfo
  {author} {\bibfnamefont {L.}~\bibnamefont {Novotny}},\ }\href@noop {}
  {\bibfield  {journal} {\bibinfo  {journal} {Physical Review Letters}\
  }\textbf {\bibinfo {volume} {114}},\ \bibinfo {pages} {163903} (\bibinfo
  {year} {2015})}\BibitemShut {NoStop}%
\bibitem [{\citenamefont {Vaskin}\ \emph {et~al.}(2019)\citenamefont {Vaskin},
  \citenamefont {Mashhadi}, \citenamefont {Steinert}, \citenamefont {Chong},
  \citenamefont {Keene}, \citenamefont {Nanz}, \citenamefont {Abass},
  \citenamefont {Rusak}, \citenamefont {Choi}, \citenamefont
  {Fernandez-Corbaton}, \citenamefont {Pertsch}, \citenamefont {Rockstuhl},
  \citenamefont {Noginov}, \citenamefont {Kivshar}, \citenamefont {Neshev},
  \citenamefont {Noginova},\ and\ \citenamefont {Staude}}]{Vaskin2018}%
  \BibitemOpen
  \bibfield  {author} {\bibinfo {author} {\bibfnamefont {A.}~\bibnamefont
  {Vaskin}}, \bibinfo {author} {\bibfnamefont {S.}~\bibnamefont {Mashhadi}},
  \bibinfo {author} {\bibfnamefont {M.}~\bibnamefont {Steinert}}, \bibinfo
  {author} {\bibfnamefont {K.~E.}\ \bibnamefont {Chong}}, \bibinfo {author}
  {\bibfnamefont {D.}~\bibnamefont {Keene}}, \bibinfo {author} {\bibfnamefont
  {S.}~\bibnamefont {Nanz}}, \bibinfo {author} {\bibfnamefont {A.}~\bibnamefont
  {Abass}}, \bibinfo {author} {\bibfnamefont {E.}~\bibnamefont {Rusak}},
  \bibinfo {author} {\bibfnamefont {D.-Y.}\ \bibnamefont {Choi}}, \bibinfo
  {author} {\bibfnamefont {I.}~\bibnamefont {Fernandez-Corbaton}}, \bibinfo
  {author} {\bibfnamefont {T.}~\bibnamefont {Pertsch}}, \bibinfo {author}
  {\bibfnamefont {C.}~\bibnamefont {Rockstuhl}}, \bibinfo {author}
  {\bibfnamefont {M.~A.}\ \bibnamefont {Noginov}}, \bibinfo {author}
  {\bibfnamefont {Y.~S.}\ \bibnamefont {Kivshar}}, \bibinfo {author}
  {\bibfnamefont {D.~N.}\ \bibnamefont {Neshev}}, \bibinfo {author}
  {\bibfnamefont {N.}~\bibnamefont {Noginova}}, \ and\ \bibinfo {author}
  {\bibfnamefont {I.}~\bibnamefont {Staude}},\ }\href@noop {} {\bibfield
  {journal} {\bibinfo  {journal} {Nano Letters}\ }\textbf {\bibinfo {volume}
  {19}},\ \bibinfo {pages} {1015} (\bibinfo {year} {2019})}\BibitemShut
  {NoStop}%
\bibitem [{\citenamefont {Kern}\ and\ \citenamefont {Martin}(2012)}]{Kern2012}%
  \BibitemOpen
  \bibfield  {author} {\bibinfo {author} {\bibfnamefont {A.}~\bibnamefont
  {Kern}}\ and\ \bibinfo {author} {\bibfnamefont {O.~J.}\ \bibnamefont
  {Martin}},\ }\href@noop {} {\bibfield  {journal} {\bibinfo  {journal}
  {Physical Review A}\ }\textbf {\bibinfo {volume} {85}},\ \bibinfo {pages}
  {022501} (\bibinfo {year} {2012})}\BibitemShut {NoStop}%
\bibitem [{\citenamefont {Filter}\ \emph {et~al.}(2012)\citenamefont {Filter},
  \citenamefont {M{\"u}hlig}, \citenamefont {Eichelkraut}, \citenamefont
  {Rockstuhl},\ and\ \citenamefont {Lederer}}]{Filter2012}%
  \BibitemOpen
  \bibfield  {author} {\bibinfo {author} {\bibfnamefont {R.}~\bibnamefont
  {Filter}}, \bibinfo {author} {\bibfnamefont {S.}~\bibnamefont {M{\"u}hlig}},
  \bibinfo {author} {\bibfnamefont {T.}~\bibnamefont {Eichelkraut}}, \bibinfo
  {author} {\bibfnamefont {C.}~\bibnamefont {Rockstuhl}}, \ and\ \bibinfo
  {author} {\bibfnamefont {F.}~\bibnamefont {Lederer}},\ }\href@noop {}
  {\bibfield  {journal} {\bibinfo  {journal} {Physical Review B}\ }\textbf
  {\bibinfo {volume} {86}},\ \bibinfo {pages} {035404} (\bibinfo {year}
  {2012})}\BibitemShut {NoStop}%
\bibitem [{\citenamefont {Yannopapas}\ and\ \citenamefont
  {Paspalakis}(2015)}]{Yannopapas2015}%
  \BibitemOpen
  \bibfield  {author} {\bibinfo {author} {\bibfnamefont {V.}~\bibnamefont
  {Yannopapas}}\ and\ \bibinfo {author} {\bibfnamefont {E.}~\bibnamefont
  {Paspalakis}},\ }\href@noop {} {\bibfield  {journal} {\bibinfo  {journal}
  {Journal of Modern Optics}\ }\textbf {\bibinfo {volume} {62}},\ \bibinfo
  {pages} {1435} (\bibinfo {year} {2015})}\BibitemShut {NoStop}%
\bibitem [{\citenamefont {Rivera}\ \emph {et~al.}(2016)\citenamefont {Rivera},
  \citenamefont {Kaminer}, \citenamefont {Zhen}, \citenamefont {Joannopoulos},\
  and\ \citenamefont {Solja{\v{c}}i{\'c}}}]{Rivera2016}%
  \BibitemOpen
  \bibfield  {author} {\bibinfo {author} {\bibfnamefont {N.}~\bibnamefont
  {Rivera}}, \bibinfo {author} {\bibfnamefont {I.}~\bibnamefont {Kaminer}},
  \bibinfo {author} {\bibfnamefont {B.}~\bibnamefont {Zhen}}, \bibinfo {author}
  {\bibfnamefont {J.~D.}\ \bibnamefont {Joannopoulos}}, \ and\ \bibinfo
  {author} {\bibfnamefont {M.}~\bibnamefont {Solja{\v{c}}i{\'c}}},\ }\href@noop
  {} {\bibfield  {journal} {\bibinfo  {journal} {Science}\ }\textbf {\bibinfo
  {volume} {353}},\ \bibinfo {pages} {263} (\bibinfo {year}
  {2016})}\BibitemShut {NoStop}%
\bibitem [{\citenamefont {Neuman}\ \emph {et~al.}(2018)\citenamefont {Neuman},
  \citenamefont {Esteban}, \citenamefont {Casanova}, \citenamefont
  {Garc{\'\i}a-Vidal},\ and\ \citenamefont {Aizpurua}}]{Neuman2018}%
  \BibitemOpen
  \bibfield  {author} {\bibinfo {author} {\bibfnamefont {T.}~\bibnamefont
  {Neuman}}, \bibinfo {author} {\bibfnamefont {R.}~\bibnamefont {Esteban}},
  \bibinfo {author} {\bibfnamefont {D.}~\bibnamefont {Casanova}}, \bibinfo
  {author} {\bibfnamefont {F.~J.}\ \bibnamefont {Garc{\'\i}a-Vidal}}, \ and\
  \bibinfo {author} {\bibfnamefont {J.}~\bibnamefont {Aizpurua}},\ }\href@noop
  {} {\bibfield  {journal} {\bibinfo  {journal} {Nano Letters}\ }\textbf
  {\bibinfo {volume} {18}},\ \bibinfo {pages} {2358} (\bibinfo {year}
  {2018})}\BibitemShut {NoStop}%
\bibitem [{\citenamefont {Tighineanu}\ \emph {et~al.}(2014)\citenamefont
  {Tighineanu}, \citenamefont {Andersen}, \citenamefont {S{\o}rensen},
  \citenamefont {Stobbe},\ and\ \citenamefont {Lodahl}}]{Tighineanu2014}%
  \BibitemOpen
  \bibfield  {author} {\bibinfo {author} {\bibfnamefont {P.}~\bibnamefont
  {Tighineanu}}, \bibinfo {author} {\bibfnamefont {M.~L.}\ \bibnamefont
  {Andersen}}, \bibinfo {author} {\bibfnamefont {A.~S.}\ \bibnamefont
  {S{\o}rensen}}, \bibinfo {author} {\bibfnamefont {S.}~\bibnamefont {Stobbe}},
  \ and\ \bibinfo {author} {\bibfnamefont {P.}~\bibnamefont {Lodahl}},\
  }\href@noop {} {\bibfield  {journal} {\bibinfo  {journal} {Physical Review
  Letters}\ }\textbf {\bibinfo {volume} {113}},\ \bibinfo {pages} {043601}
  (\bibinfo {year} {2014})}\BibitemShut {NoStop}%
\bibitem [{\citenamefont {Li}\ \emph {et~al.}(2018)\citenamefont {Li},
  \citenamefont {Karaveli}, \citenamefont {Cueff}, \citenamefont {Li},\ and\
  \citenamefont {Zia}}]{Li2018}%
  \BibitemOpen
  \bibfield  {author} {\bibinfo {author} {\bibfnamefont {D.}~\bibnamefont
  {Li}}, \bibinfo {author} {\bibfnamefont {S.}~\bibnamefont {Karaveli}},
  \bibinfo {author} {\bibfnamefont {S.}~\bibnamefont {Cueff}}, \bibinfo
  {author} {\bibfnamefont {W.}~\bibnamefont {Li}}, \ and\ \bibinfo {author}
  {\bibfnamefont {R.}~\bibnamefont {Zia}},\ }\href@noop {} {\bibfield
  {journal} {\bibinfo  {journal} {Physical Review Letters}\ }\textbf {\bibinfo
  {volume} {121}},\ \bibinfo {pages} {227403} (\bibinfo {year}
  {2018})}\BibitemShut {NoStop}%
\bibitem [{\citenamefont {Rusak}\ \emph {et~al.}(2019)\citenamefont {Rusak},
  \citenamefont {Straubel}, \citenamefont {G{\l}adysz}, \citenamefont
  {G{\"o}ddel}, \citenamefont {K{\k{e}}dziorski}, \citenamefont {K{\"u}hn},
  \citenamefont {Weigend}, \citenamefont {Rockstuhl},\ and\ \citenamefont
  {S{\l}owik}}]{Rusak2019}%
  \BibitemOpen
  \bibfield  {author} {\bibinfo {author} {\bibfnamefont {E.}~\bibnamefont
  {Rusak}}, \bibinfo {author} {\bibfnamefont {J.}~\bibnamefont {Straubel}},
  \bibinfo {author} {\bibfnamefont {P.}~\bibnamefont {G{\l}adysz}}, \bibinfo
  {author} {\bibfnamefont {M.}~\bibnamefont {G{\"o}ddel}}, \bibinfo {author}
  {\bibfnamefont {A.}~\bibnamefont {K{\k{e}}dziorski}}, \bibinfo {author}
  {\bibfnamefont {M.}~\bibnamefont {K{\"u}hn}}, \bibinfo {author}
  {\bibfnamefont {F.}~\bibnamefont {Weigend}}, \bibinfo {author} {\bibfnamefont
  {C.}~\bibnamefont {Rockstuhl}}, \ and\ \bibinfo {author} {\bibfnamefont
  {K.}~\bibnamefont {S{\l}owik}},\ }\href@noop {} {\bibfield  {journal}
  {\bibinfo  {journal} {arXiv preprint arXiv:1905.08482}\ } (\bibinfo {year}
  {2019})}\BibitemShut {NoStop}%
\bibitem [{\citenamefont {Barnett}\ \emph {et~al.}(1996)\citenamefont
  {Barnett}, \citenamefont {Huttner}, \citenamefont {Loudon},\ and\
  \citenamefont {Matloob}}]{Barnett1996}%
  \BibitemOpen
  \bibfield  {author} {\bibinfo {author} {\bibfnamefont {S.~M.}\ \bibnamefont
  {Barnett}}, \bibinfo {author} {\bibfnamefont {B.}~\bibnamefont {Huttner}},
  \bibinfo {author} {\bibfnamefont {R.}~\bibnamefont {Loudon}}, \ and\ \bibinfo
  {author} {\bibfnamefont {R.}~\bibnamefont {Matloob}},\ }\href@noop {}
  {\bibfield  {journal} {\bibinfo  {journal} {Journal of Physics B: Atomic,
  Molecular and Optical Physics}\ }\textbf {\bibinfo {volume} {29}},\ \bibinfo
  {pages} {3763} (\bibinfo {year} {1996})}\BibitemShut {NoStop}%
\bibitem [{\citenamefont {Barron}\ and\ \citenamefont {Gray}(1973)}]{barron}%
  \BibitemOpen
  \bibfield  {author} {\bibinfo {author} {\bibfnamefont {L.~D.}\ \bibnamefont
  {Barron}}\ and\ \bibinfo {author} {\bibfnamefont {C.~G.}\ \bibnamefont
  {Gray}},\ }\href {http://stacks.iop.org/0301-0015/6/i=1/a=006} {\bibfield
  {journal} {\bibinfo  {journal} {Journal of Physics A: Mathematical, Nuclear
  and General}\ }\textbf {\bibinfo {volume} {6}},\ \bibinfo {pages} {59}
  (\bibinfo {year} {1973})}\BibitemShut {NoStop}%
\bibitem [{\citenamefont {Huttner}\ and\ \citenamefont
  {Barnett}(1992)}]{Huttner1992}%
  \BibitemOpen
  \bibfield  {author} {\bibinfo {author} {\bibfnamefont {B.}~\bibnamefont
  {Huttner}}\ and\ \bibinfo {author} {\bibfnamefont {S.~M.}\ \bibnamefont
  {Barnett}},\ }\href@noop {} {\bibfield  {journal} {\bibinfo  {journal}
  {Physical Review A}\ }\textbf {\bibinfo {volume} {46}},\ \bibinfo {pages}
  {4306} (\bibinfo {year} {1992})}\BibitemShut {NoStop}%
\bibitem [{\citenamefont {Matloob}\ \emph {et~al.}(1995)\citenamefont
  {Matloob}, \citenamefont {Loudon}, \citenamefont {Barnett},\ and\
  \citenamefont {Jeffers}}]{Matloob1995}%
  \BibitemOpen
  \bibfield  {author} {\bibinfo {author} {\bibfnamefont {R.}~\bibnamefont
  {Matloob}}, \bibinfo {author} {\bibfnamefont {R.}~\bibnamefont {Loudon}},
  \bibinfo {author} {\bibfnamefont {S.~M.}\ \bibnamefont {Barnett}}, \ and\
  \bibinfo {author} {\bibfnamefont {J.}~\bibnamefont {Jeffers}},\ }\href@noop
  {} {\bibfield  {journal} {\bibinfo  {journal} {Physical Review A}\ }\textbf
  {\bibinfo {volume} {52}},\ \bibinfo {pages} {4823} (\bibinfo {year}
  {1995})}\BibitemShut {NoStop}%
\bibitem [{\citenamefont {Gruner}\ and\ \citenamefont
  {Welsch}(1996)}]{Gruner1996}%
  \BibitemOpen
  \bibfield  {author} {\bibinfo {author} {\bibfnamefont {T.}~\bibnamefont
  {Gruner}}\ and\ \bibinfo {author} {\bibfnamefont {D.-G.}\ \bibnamefont
  {Welsch}},\ }\href@noop {} {\bibfield  {journal} {\bibinfo  {journal}
  {Physical Review A}\ }\textbf {\bibinfo {volume} {53}},\ \bibinfo {pages}
  {1818} (\bibinfo {year} {1996})}\BibitemShut {NoStop}%
\bibitem [{\citenamefont {Dung}\ \emph {et~al.}(1998)\citenamefont {Dung},
  \citenamefont {Kn{\"o}ll},\ and\ \citenamefont {Welsch}}]{Dung1998}%
  \BibitemOpen
  \bibfield  {author} {\bibinfo {author} {\bibfnamefont {H.~T.}\ \bibnamefont
  {Dung}}, \bibinfo {author} {\bibfnamefont {L.}~\bibnamefont {Kn{\"o}ll}}, \
  and\ \bibinfo {author} {\bibfnamefont {D.-G.}\ \bibnamefont {Welsch}},\
  }\href@noop {} {\bibfield  {journal} {\bibinfo  {journal} {Physical Review
  A}\ }\textbf {\bibinfo {volume} {57}},\ \bibinfo {pages} {3931} (\bibinfo
  {year} {1998})}\BibitemShut {NoStop}%
\bibitem [{\citenamefont {Novotny}\ and\ \citenamefont
  {Hecht}(2006)}]{novotny}%
  \BibitemOpen
  \bibfield  {author} {\bibinfo {author} {\bibfnamefont {L.}~\bibnamefont
  {Novotny}}\ and\ \bibinfo {author} {\bibfnamefont {B.}~\bibnamefont
  {Hecht}},\ }\href {\doibase 10.1017/CBO9780511813535} {\emph {\bibinfo
  {title} {Principles of Nano-Optics}}}\ (\bibinfo  {publisher} {Cambridge
  University Press},\ \bibinfo {year} {2006})\BibitemShut {NoStop}%
\bibitem [{\citenamefont {Dung}\ \emph {et~al.}(2000)\citenamefont {Dung},
  \citenamefont {Kn{\"o}ll},\ and\ \citenamefont {Welsch}}]{Dung2000}%
  \BibitemOpen
  \bibfield  {author} {\bibinfo {author} {\bibfnamefont {H.~T.}\ \bibnamefont
  {Dung}}, \bibinfo {author} {\bibfnamefont {L.}~\bibnamefont {Kn{\"o}ll}}, \
  and\ \bibinfo {author} {\bibfnamefont {D.-G.}\ \bibnamefont {Welsch}},\
  }\href@noop {} {\bibfield  {journal} {\bibinfo  {journal} {Physical Review
  A}\ }\textbf {\bibinfo {volume} {62}},\ \bibinfo {pages} {053804} (\bibinfo
  {year} {2000})}\BibitemShut {NoStop}%
\bibitem [{\citenamefont {Martin}\ and\ \citenamefont
  {Piller}(1998)}]{Martin1998}%
  \BibitemOpen
  \bibfield  {author} {\bibinfo {author} {\bibfnamefont {O.~J.}\ \bibnamefont
  {Martin}}\ and\ \bibinfo {author} {\bibfnamefont {N.~B.}\ \bibnamefont
  {Piller}},\ }\href@noop {} {\bibfield  {journal} {\bibinfo  {journal}
  {Physical Review E}\ }\textbf {\bibinfo {volume} {58}},\ \bibinfo {pages}
  {3909} (\bibinfo {year} {1998})}\BibitemShut {NoStop}%
\bibitem [{\citenamefont {Sinha}\ \emph {et~al.}(2018)\citenamefont {Sinha},
  \citenamefont {Venkatesh},\ and\ \citenamefont {Meystre}}]{Sinha2018}%
  \BibitemOpen
  \bibfield  {author} {\bibinfo {author} {\bibfnamefont {K.}~\bibnamefont
  {Sinha}}, \bibinfo {author} {\bibfnamefont {B.~P.}\ \bibnamefont
  {Venkatesh}}, \ and\ \bibinfo {author} {\bibfnamefont {P.}~\bibnamefont
  {Meystre}},\ }\href@noop {} {\bibfield  {journal} {\bibinfo  {journal}
  {Physical Review Letters}\ }\textbf {\bibinfo {volume} {121}},\ \bibinfo
  {pages} {183605} (\bibinfo {year} {2018})}\BibitemShut {NoStop}%
\bibitem [{\citenamefont {Lukosz}\ and\ \citenamefont
  {Kunz}(1977)}]{Lukosz1977}%
  \BibitemOpen
  \bibfield  {author} {\bibinfo {author} {\bibfnamefont {W.}~\bibnamefont
  {Lukosz}}\ and\ \bibinfo {author} {\bibfnamefont {R.}~\bibnamefont {Kunz}},\
  }\href@noop {} {\bibfield  {journal} {\bibinfo  {journal} {JOSA}\ }\textbf
  {\bibinfo {volume} {67}},\ \bibinfo {pages} {1607} (\bibinfo {year}
  {1977})}\BibitemShut {NoStop}%
\bibitem [{\citenamefont {Henderson}\ and\ \citenamefont
  {Imbusch}(2006)}]{Henderson2006}%
  \BibitemOpen
  \bibfield  {author} {\bibinfo {author} {\bibfnamefont {B.}~\bibnamefont
  {Henderson}}\ and\ \bibinfo {author} {\bibfnamefont {G.~F.}\ \bibnamefont
  {Imbusch}},\ }\href@noop {} {\emph {\bibinfo {title} {Optical spectroscopy of
  inorganic solids}}},\ Vol.~\bibinfo {volume} {44}\ (\bibinfo  {publisher}
  {Oxford University Press},\ \bibinfo {year} {2006})\BibitemShut {NoStop}%
\bibitem [{\citenamefont {Craig}\ and\ \citenamefont
  {Thirunamachandran}(1998)}]{Craig1998}%
  \BibitemOpen
  \bibfield  {author} {\bibinfo {author} {\bibfnamefont {D.~P.}\ \bibnamefont
  {Craig}}\ and\ \bibinfo {author} {\bibfnamefont {T.}~\bibnamefont
  {Thirunamachandran}},\ }\href@noop {} {\emph {\bibinfo {title} {Molecular
  quantum electrodynamics: an introduction to radiation-molecule
  interactions}}}\ (\bibinfo  {publisher} {Courier Corporation},\ \bibinfo
  {year} {1998})\BibitemShut {NoStop}%
\bibitem [{\citenamefont {Hohenester}\ and\ \citenamefont
  {Trügler}(2012)}]{mnpbem}%
  \BibitemOpen
  \bibfield  {author} {\bibinfo {author} {\bibfnamefont {U.}~\bibnamefont
  {Hohenester}}\ and\ \bibinfo {author} {\bibfnamefont {A.}~\bibnamefont
  {Trügler}},\ }\href
  {http://www.sciencedirect.com/science/article/pii/S0010465511003274}
  {\bibfield  {journal} {\bibinfo  {journal} {Computer Physics Communications}\
  }\textbf {\bibinfo {volume} {183}},\ \bibinfo {pages} {370 } (\bibinfo {year}
  {2012})}\BibitemShut {NoStop}%
\bibitem [{\citenamefont {De~Abajo}\ and\ \citenamefont
  {Howie}(2002)}]{deAbajo2002}%
  \BibitemOpen
  \bibfield  {author} {\bibinfo {author} {\bibfnamefont {F.~G.}\ \bibnamefont
  {De~Abajo}}\ and\ \bibinfo {author} {\bibfnamefont {A.}~\bibnamefont
  {Howie}},\ }\href@noop {} {\bibfield  {journal} {\bibinfo  {journal}
  {Physical Review B}\ }\textbf {\bibinfo {volume} {65}},\ \bibinfo {pages}
  {115418} (\bibinfo {year} {2002})}\BibitemShut {NoStop}%
\bibitem [{\citenamefont {Hohenester}\ and\ \citenamefont
  {Trugler}(2008)}]{Hohenester2008}%
  \BibitemOpen
  \bibfield  {author} {\bibinfo {author} {\bibfnamefont {U.}~\bibnamefont
  {Hohenester}}\ and\ \bibinfo {author} {\bibfnamefont {A.}~\bibnamefont
  {Trugler}},\ }\href@noop {} {\bibfield  {journal} {\bibinfo  {journal} {IEEE
  Journal of Selected Topics in Quantum Electronics}\ }\textbf {\bibinfo
  {volume} {14}},\ \bibinfo {pages} {1430} (\bibinfo {year}
  {2008})}\BibitemShut {NoStop}%
\bibitem [{\citenamefont {De~Abajo}(2010)}]{deAbajo2010}%
  \BibitemOpen
  \bibfield  {author} {\bibinfo {author} {\bibfnamefont {F.~G.}\ \bibnamefont
  {De~Abajo}},\ }\href@noop {} {\bibfield  {journal} {\bibinfo  {journal}
  {Reviews of Modern Physics}\ }\textbf {\bibinfo {volume} {82}},\ \bibinfo
  {pages} {209} (\bibinfo {year} {2010})}\BibitemShut {NoStop}%
\bibitem [{\citenamefont {Johnson}\ and\ \citenamefont
  {Christy}(1972)}]{johnson}%
  \BibitemOpen
  \bibfield  {author} {\bibinfo {author} {\bibfnamefont {P.~B.}\ \bibnamefont
  {Johnson}}\ and\ \bibinfo {author} {\bibfnamefont {R.~W.}\ \bibnamefont
  {Christy}},\ }\href {https://link.aps.org/doi/10.1103/PhysRevB.6.4370}
  {\bibfield  {journal} {\bibinfo  {journal} {Phys. Rev. B}\ }\textbf {\bibinfo
  {volume} {6}},\ \bibinfo {pages} {4370} (\bibinfo {year} {1972})}\BibitemShut
  {NoStop}%
\bibitem [{\citenamefont {Blanchard}\ and\ \citenamefont
  {Br{\"u}ning}(2003)}]{Blanchard2003}%
  \BibitemOpen
  \bibfield  {author} {\bibinfo {author} {\bibfnamefont {P.}~\bibnamefont
  {Blanchard}}\ and\ \bibinfo {author} {\bibfnamefont {E.}~\bibnamefont
  {Br{\"u}ning}},\ }\href@noop {} {\emph {\bibinfo {title} {Mathematical
  Methods in Physics: Distributions, Hilbert Space Operators, and Variational
  Methods}}},\ Progress in Mathematical Physics\ (\bibinfo  {publisher}
  {Birkh{\"a}user},\ \bibinfo {year} {2003})\BibitemShut {NoStop}%
\bibitem [{\citenamefont {Vogel}\ and\ \citenamefont {Welsch}(2006)}]{vogel}%
  \BibitemOpen
  \bibfield  {author} {\bibinfo {author} {\bibfnamefont {W.}~\bibnamefont
  {Vogel}}\ and\ \bibinfo {author} {\bibfnamefont {D.-G.}\ \bibnamefont
  {Welsch}},\ }\href@noop {} {\emph {\bibinfo {title} {{Quantum Optics}}}}\
  (\bibinfo  {publisher} {Wiley-VCH},\ \bibinfo {year} {2006})\BibitemShut
  {NoStop}%
\bibitem [{\citenamefont {Dzsotjan}\ \emph {et~al.}(2011)\citenamefont
  {Dzsotjan}, \citenamefont {K{\"a}stel},\ and\ \citenamefont
  {Fleischhauer}}]{Dzsotjan2011}%
  \BibitemOpen
  \bibfield  {author} {\bibinfo {author} {\bibfnamefont {D.}~\bibnamefont
  {Dzsotjan}}, \bibinfo {author} {\bibfnamefont {J.}~\bibnamefont
  {K{\"a}stel}}, \ and\ \bibinfo {author} {\bibfnamefont {M.}~\bibnamefont
  {Fleischhauer}},\ }\href@noop {} {\bibfield  {journal} {\bibinfo  {journal}
  {Physical Review B}\ }\textbf {\bibinfo {volume} {84}},\ \bibinfo {pages}
  {075419} (\bibinfo {year} {2011})}\BibitemShut {NoStop}%
\end{thebibliography}%

\appendix
\section{Derivation of the effective atomic dynamics for single atomic system}\label{app:single_system}

We begin with the Heisenberg equations for both atomic systems and fields $i\hbar\dot{O}=\left[O,\mathcal{H}\right]$ where $O$ is an arbitrary operator and $\mathcal{H}=\mathcal{H}_0+\mathcal{H}_\mathrm{field}+\mathcal{H}_\mathrm{int}$ is the total Hamiltonian.
The equations read
\begin{widetext}
\begin{eqnarray}
\label{eq:sigma}
\dot{\sigma} &=& -i\omega_0\sigma  - \frac{i}{\hbar}\sigma_z 
\left(
\mathbf{d}^\dagger\cdot\mathbf{E}^{(+)}_0 + 
\mathbf{m}^\dagger\cdot\mathbf{B}^{(+)}_0+ 
\mathbf{Q}^\dagger:\nabla\mathbf{E}^{(+)}_0\right)\\
\label{eq:sigmaz}
\dot{\sigma}_z &=& \frac{2i}{\hbar} \left\{ \sigma^{\dagger}\left(\mathbf{d}^{\dagger}\cdot\mathbf{E}^{(+)}_0+\mathbf{m}^\dagger\cdot\mathbf{B}^{(+)}_0+\mathbf{Q}^{\dagger}:\nabla\mathbf{E}^{(+)}_0\right)-\left(\mathbf{E}^{(-)}_0\cdot\mathbf{d}+\mathbf{B}^{(-)}_0\cdot\mathbf{m}+\nabla\mathbf{E}^{(-)}_0:\mathbf{Q}\right)\sigma\right\}\\
\label{eq:f_oper}
\dot{f}_{\omega,j} &=& -i\omega f_{\omega,j} 
+\frac{i}{\hbar}\left[f_{\omega,j},\mathbf{E}^{(-)}_0\cdot\mathbf{d}+\mathbf{B}^{(-)}_0\cdot\mathbf{m}+\nabla\mathbf{E}^{(-)}_0:\mathbf{Q}\right]\sigma \nonumber\\
&=&-i\omega f_{\omega,j}\left(\mathbf{r}\right)+\sqrt{\frac{\mathrm{Im}\,\varepsilon\left(\mathbf{r},\omega\right)}{\hbar\pi\epsilon_{0}}}\frac{\omega^{2}}{c^{2}}\sum_{m}D^{r^{\prime\prime}}_m(\omega)G_{mj}^{\star}\left(\mathbf{r}^{\prime\prime},\mathbf{r},\omega\right)|_{\mathbf{r}^{\prime\prime}=\mathbf{r}_{0}}\sigma,
\end{eqnarray}
where we have used the commutation relations of atomic Pauli operators and of bosonic field operators given by Eqs.~(\ref{eq:f_commutators}) of the main text. We have used the representation of double dot product 
$\nabla\mathbf{E}^{(-)}\left(\mathbf{r}\right):\mathbf{Q}=\sum_{ij}Q_{ji}\partial_iE_j^{(-)}\left(\mathbf{r}\right)$ and of rotation $\left[\nabla \times \mathbf{E}^{(-)}\left(\mathbf{r}\right)\right]_i =  
\sum_{jk}\epsilon_{ijk} \partial_{r_j} {E}_k^{(-)}\left(\mathbf{r}\right)$.
We now formally integrate Eq.~(\ref{eq:f_oper}) to obtain
\begin{equation}\label{eq:field_integrated}
f_{\omega,j}(t) =f_{\omega,j}^{\mathrm{free}}\left(\mathbf{r}\right)
+\sqrt{\frac{\mathrm{Im}\,\varepsilon\left(\mathbf{r},\omega\right)}{\hbar\pi\epsilon_{0}}}\frac{\omega^{2}}{c^{2}}\left\{ \sum_{m}D^{r^{\prime\prime}}_m(\omega)G_{mj}^{\star}\left(\mathbf{r}^{\prime\prime},\mathbf{r},\omega\right)|_{\mathbf{r}^{\prime\prime}=\mathbf{r}_{0}}\right\} \int_{0}^{t}dt^{\prime}\sigma\left(t^{\prime}\right)e^{-i\omega (t-t^{\prime})}
\end{equation}
\end{widetext}
where 
$f^\mathrm{free}_{\omega,j}(t)$ describes the free field, and the integral accounts for the influence of the atomic system on the field. Since we focus on the atomic dynamics, the above result is inserted in the place of fields in Eqs.~(\ref{eq:sigma}-\ref{eq:sigmaz}). The resulting integro-differential equations can be simplified if the coupling with the environment is weak:
In the free-atomic-system case, the operator
$\sigma(t)=\sigma(0) e^{-i\omega_0 t}$ evolves freely.
The photonic environment, which represents vacuum fluctuations, is characterized by a set of modes continuously distributed in frequencies and weakly coupled to the atomic system. 
Such environment introduces
small modifications to the atomic evolution, which can be quantified in terms of a slowly varying envelope $\tilde{\sigma}(t)$ modulated upon the free oscillations: $\sigma(t)=\tilde{\sigma}(t)e^{-i\omega_0t}$. The assumption in the Markovian approximation is that the envelope changes little over the time interval $\omega_0^{-1}$ around $t^\prime \approx t$, where the oscillating term takes significant values. 
This yields
\begin{eqnarray}\label{eq:markovian_approximation}
&&\int_{0}^{t}dt^{\prime}\sigma\left(t^{\prime}\right)e^{-i\omega (t-t^{\prime})}=
e^{-i\omega_0 t}\int_{0}^{t}dt^{\prime}\tilde{\sigma}\left(t^{\prime}\right)e^{-i(\omega-\omega_0)(t-t^{\prime})}\nonumber\\
&&\approx \sigma(t) \int_0^\infty  e^{i(\omega-\omega_0)\tau} \, d\tau  \nonumber\\
&&=\sigma(t)\left\{\pi\delta(\omega-\omega_0)+i\mathcal{P}\left[(\omega-\omega_0)^{-1}\right]\right\}. 
\end{eqnarray}
Due to this time-scale mismatch, the time of interest $t\gg\omega_0^{-1}$ and we could replace the upper limit of the above integral with infinity. In the last step we have used the Sokhotski–Plemelj theorem \cite{Blanchard2003}. The symbol $\mathcal{P}$ denotes the Cauchy principal value.
The interpretation is that the evolution of the system is affected only by its present state, and memory effects are negligible. 

Please note that here and in the following part of the derivation it is crucial to consequently keep normal operator ordering, as it naturally follows from the Hamiltonian in Eq.~(\ref{eq:hamiltonian}) of the main text. This is because the Markovian approximation in general affects the commutation relations, and atomic and field operators are no longer independent. 

Having inserted Eq.~(\ref{eq:field_integrated}) into Eqs.~(\ref{eq:sigma}) and (\ref{eq:sigmaz}) and applying the Markovian approximation given by Eq.~(\ref{eq:markovian_approximation}), we arrive into Eqs.~(\ref{eq:sigma_dynamics}) and (\ref{eq:sigmaz_dynamics}) of the main text. We have additionally used the Green's tensor property \cite{vogel}
\begin{multline}
\label{eq:G_property}
\frac{\omega^2}{c^2}\int d^3r \mathrm{Im}\,\epsilon \left(\mathbf{r},\omega\right)
\sum_nG_\mathrm{mn}\left( \mathbf{r}_1, \mathbf{r}, \omega\right)
G^\star_\mathrm{pn}\left( \mathbf{r}_2, \mathbf{r}, \omega\right)=\\
= \mathrm{Im}\,G_\mathrm{mp}\left( \mathbf{r}_1, \mathbf{r}_2, \omega\right),
\end{multline}
and the reciprocity relation
\begin{equation}\label{eq:G_reciprocity}
G_{ij}\left(\mathbf{r}^\prime,\mathbf{r},\omega\right)=G_{ji}\left(\mathbf{r},\mathbf{r}^\prime,\omega\right).
\end{equation}

\section{Generalization to the case of multiple atomic systems}\label{app:multiple_systems}
We begin with the Hamiltonian given by Eq.~(\ref{eq:hamiltonian_multi}) and repeat steps corresponding to Eqs.~(\ref{eq:sigma}-\ref{eq:field_integrated}). To perform the Markovian approximation, we note that we allow few atomic species to oscillate with in general different frequencies $\omega_\alpha$. We assume that these frequencies are relatively close to each other $\Delta_\omega \equiv \max_{\alpha,\alpha^\prime}|\omega_\alpha-\omega_\alpha^\prime|\ll \bar{\omega}$, where $\bar{\omega}$ is the average transition frequency. In this case, in the joint response of the atomic systems we expect beating: fast oscillations with average frequency $\bar{\omega}$ are modulated with an envelope varying at timescales of the order of $\Delta_\omega^{-1}$. Additionally, we assume that the Green's tensor has a relatively broad peak at the frequency range of interest, covering all atomic frequencies, and assume that nevertheless the two-level approximation for each system still holds.  
We choose a corresponding ansatz to represent atomic operators for each $\alpha$: $\sigma_\alpha(t) = \tilde{\sigma}_\alpha(t) e^{-i\bar{\omega} t}$, i.e. we separate the fast oscillation and include the slowly varying envelope in $\tilde{\sigma}_\alpha(t)$. With this ansatz we can rewrite Eq.~(\ref{eq:markovian_approximation}) in the multi-atomic-system case. As a result, using also Eqs.~(\ref{eq:G_property},\ref{eq:G_reciprocity}), we arrive at Eqs.~(\ref{eq:sigma_multiatom} \& \ref{eq:sigmaz_multiatom}) of the main text.
Effective coupling strengths $\xi_{\alpha,\beta}$ and collective decay rates $\gamma_{\alpha.\beta}$ are derived as the result, expressed through the following operators
\begin{widetext}
\begin{eqnarray}
R_{mn}\left(\omega\right) &=& \frac{1}{\hbar\pi\epsilon_{0}c^{2}}\left\{
\mathrm{Re}\left[d_m^\star d_n\right]
+\mathrm{Re}\left[d_m^\star \sum_l\left(Q_{nl}+i\omega^{-1}\sum_s\epsilon_{sln}m_s\right)\right]\partial_{r_l}
+\mathrm{Re}\left[\sum_k\left(Q_{mk}^\star-i\omega^{-1}\sum_p\epsilon_{pkm}m_p^\star\right)d_n\right]\partial_{r_k^\prime}\right.\nonumber\\
&&+\left.\mathrm{Re}\left[d_m^\star \sum_l\left(Q_{nl}+i\omega^{-1}\sum_s\epsilon_{sln}m_s\right)\right]\left[\sum_k\left(Q_{mk}^\star-i\omega^{-1}\sum_p\epsilon_{pkm}m_p^\star\right)d_n\right]\partial_{r_k^\prime}\partial_{r_l}\right\} \nonumber\\
&\equiv & \frac{1}{\hbar\pi\epsilon_{0}c^{2}}\mathrm{Re} \left[{D_m^{r^\prime}}^\dagger(\omega)D_n^r(\omega) \right]\label{eq:Rmn}\\ 
I_{mn}\left(\omega\right) &=& \frac{1}{\hbar\pi\epsilon_{0}c^{2}}\left\{
\mathrm{Im}\left[d_m^\star d_n\right]
+\mathrm{Im}\left[d_m^\star \sum_l\left(Q_{nl}+i\omega^{-1}\sum_s\epsilon_{sln}m_s\right)\right]\partial_{r_l}
+\mathrm{Im}\left[\sum_k\left(Q_{mk}^\star-i\omega^{-1}\sum_p\epsilon_{pkm}m_p^\star\right)d_n\right]\partial_{r_k^\prime}\right.\nonumber\\
&&+\left.\mathrm{Im}\left[d_m^\star \sum_l\left(Q_{nl}+i\omega^{-1}\sum_s\epsilon_{sln}m_s\right)\right]\left[\sum_k\left(Q_{mk}^\star-i\omega^{-1}\sum_p\epsilon_{pkm}m_p^\star\right)d_n\right]\partial_{r_k^\prime}\partial_{r_l}
\right\}\nonumber \\
&\equiv &\frac{1}{\hbar\pi\epsilon_{0}c^{2}}\mathrm{Im} \left[{D_m^{r^\prime}}^\dagger(\omega)D_n^r(\omega)\right].\label{eq:Imn}
\end{eqnarray}
\end{widetext}

\section{Replacing principal value integrals}\label{app:kramers}
For purposes of this section, we note that the real and imaginary components of the product of generalized multipole moments, given by Eqs.~(\ref{eq:Rmn} \& \ref{eq:Imn}) of Appendix \ref{app:multiple_systems}, have the functional dependence on frequency of the form  $R_{mn}(\omega),I_{mn}(\omega)=f_0+\frac{1}{\omega}f_1+\frac{1}{\omega^{2}}f_2$, where $f_{0,1,2}$ are frequency-independent, real-valued parameters possibly multiplied by spatial differentiation operators to act on Green's tensors. 
As mentioned before, we assume that the derivatives exist at positions of atomic systems, i.e. these system should not be placed exactly at interfaces of different media. Then, differentiation over spatial coordinates and integration over frequency are interchangeable. With this assumption, the dependence on spatial coordinates plays no further role throughout this section, therefore we use the simplified notation $G_{mn}\left(\mathbf{r},\mathbf{r}^{\prime},\omega\right) \equiv G_{mn}\left(\omega\right)$. We follow and generalize considerations in Ref.~\cite{Dzsotjan2011} to eliminate principle-value integrals from terms describing energy shifts. 

Equations (\ref{eq:hamiltonian_effective} \& \ref{eq:gamma_collective}) of the main text are expressed through principal-value integrals of the form
\begin{eqnarray}
    &&I = \mathcal{P}\int_{0}^{\infty}d\omega\frac{\omega^{2}}{\omega-\omega_{0}}\left(f_0+\frac{1}{\omega}f_1+\frac{1}{\omega^{2}}f_2\right)\mathrm{Im}\,G_{mn}\left(\omega\right)  \nonumber\\
    &&=\underbrace{\mathcal{P}\int_{0}^{\infty}d\omega\frac{\omega^{3}}{\omega^{2}-\omega_{0}^{2}}\left(f_0+\frac{1}{\omega}f_1+\frac{1}{\omega^{2}}f_2\right)\mathrm{Im}\,G_{mn}\left(\omega\right)}_{I_{1}}\label{eq:I1_I2}\\ &&+\underbrace{\mathcal{P}\int_{0}^{\infty}d\omega\frac{\omega^{2}}{\omega^{2}-\omega_{0}^{2}}\omega_{0}\left(f_0+\frac{1}{\omega}f_1+\frac{1}{\omega^{2}}f_2\right)\mathrm{Im}\,G_{mn}\left(\omega\right)}_{I_{2}} \nonumber
\end{eqnarray}
From Kramers-Kronig relations we have 
\begin{eqnarray}
&&\omega_{0}^{2}\mathrm{Re}\,G_{mn}\left(\omega_{0}\right)	=\frac{\omega_{0}^{2}}{\pi}\mathcal{P}\int_{-\infty}^{\infty}d\omega\frac{1}{\omega-\omega_{0}}\mathrm{Im}\,G_{mn}\left(\omega\right) \nonumber \\
&&=\frac{2\omega_{0}^{2}}{\pi}\mathcal{P}\int_{0}^{\infty}d\omega\frac{\omega}{\omega^{2}-\omega_{0}^{2}}\mathrm{Im}\,G_{mn}\left(\omega\right)\label{eq:G_is_real}\\
&&=\frac{2}{\pi}\mathcal{P}\int_{0}^{\infty}d\omega\frac{\omega^{3}}{\omega^{2}-\omega_{0}^{2}}\mathrm{Im}\,G_{mn}\left(\omega\right), \label{eq:large_freq}
\end{eqnarray}
where relation (\ref{eq:G_is_real}) follows from the assumption of real-valued Green's propagator in time domain $\mathbf{G}\left(-\omega^{\star}\right)=\mathbf{G}^{\star}\left(\omega\right)$, while Eq.~(\ref{eq:large_freq}) is justified for large frequencies, for which the peak around $\omega_0$ is shifted sufficiently far away from $0$ \cite{Dzsotjan2011}. The resulting expression allows us to simplify the first term in Eq.~(\ref{eq:I1_I2}) to the form
\begin{eqnarray}
I_1 &=& \frac{\pi}{2}\omega_{0}^{2}\left(f_0+\frac{1}{\omega}f_1+\frac{1}{\omega^{2}}f_2\right)G_{mn}^{\prime}\left(\omega_{0}\right).
\end{eqnarray}
\begin{figure}
    \centering
    \includegraphics[width=8cm]{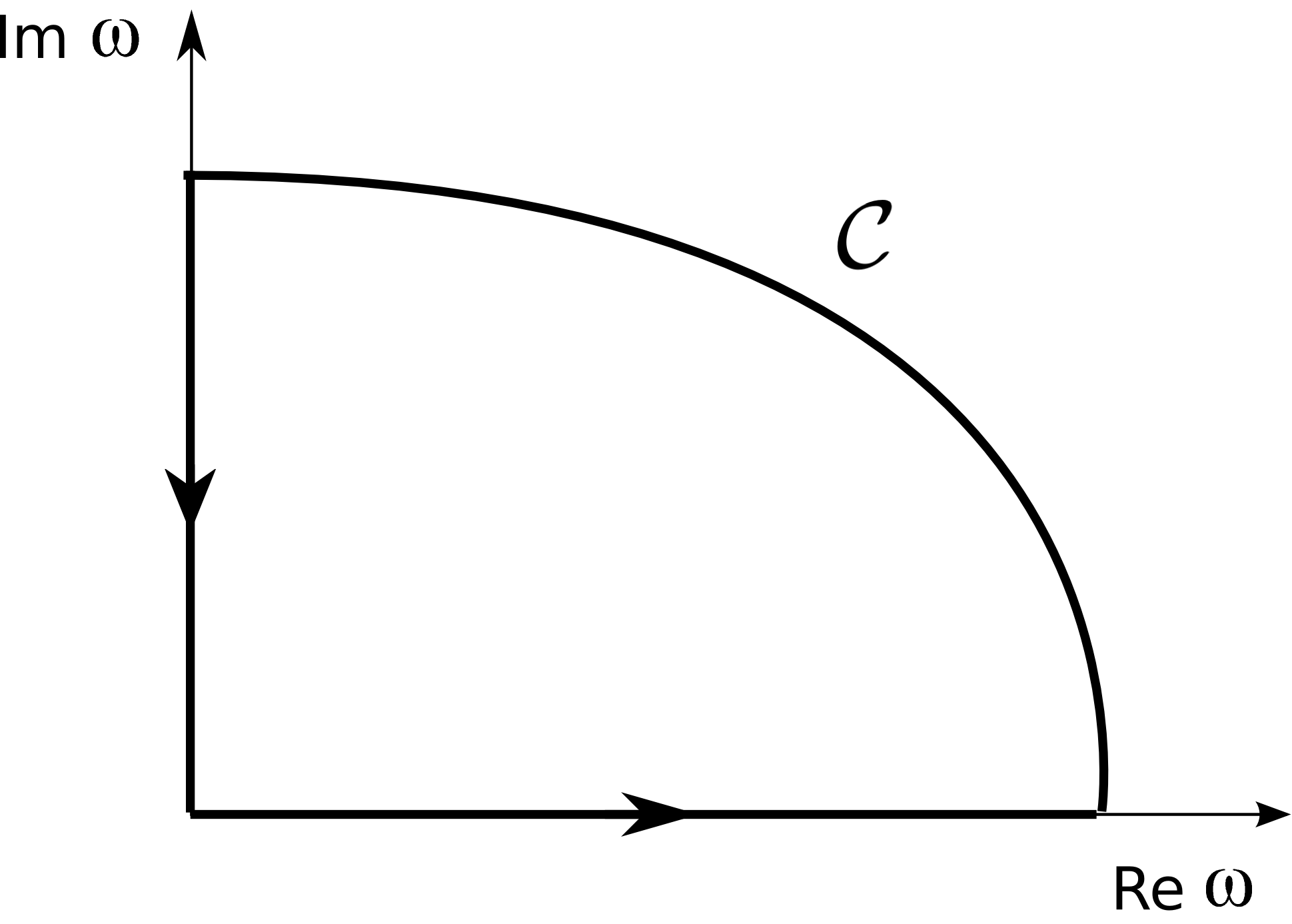}
    \caption{Contour for complex-plane integration.}
    \label{fig:contour}
\end{figure}
The second term in Eq.~(\ref{eq:I1_I2}) can be rewritten as
\begin{equation}
    I_2=\mathrm{Im}\left\{ \mathcal{P}\int_{0}^{\infty}d\omega\frac{\omega^{2}}{\omega^{2}-\omega_{0}^{2}}\omega_{0}\left[f_0+\frac{1}{\omega}f_1+\frac{1}{\omega^{2}}f_2\right]G_{mn}\left(\omega\right)\right\}.
\end{equation}
The principal value above can be resolved as
\begin{eqnarray}
\mathcal{P}\frac{\omega_{0}}{\omega^{2}-\omega_{0}^{2}}	=\frac{1}{2}\lim_{\epsilon\rightarrow0}\left(\frac{1}{\omega-\omega_{0}-i\epsilon}-i\pi\delta\left(\omega-\omega_{0}\right)\right.\nonumber\\
-\left.\frac{1}{\omega+\omega_{0}-i\epsilon}+i\pi\delta\left(\omega+\omega_{0}\right)\right),
\end{eqnarray}
which yields
\begin{widetext}
\begin{eqnarray}
    I_2 &=& \mathrm{Im}\left\{ \int_{0}^{\infty}d\omega\frac{1}{2}\lim_{\epsilon\rightarrow0}\left(\frac{1}{\omega-\omega_{0}-i\epsilon}-\frac{1}{\omega+\omega_{0}-i\epsilon}\right)\omega^{2}\left[f_0+\frac{1}{\omega}f_1+\frac{1}{\omega^{2}}f_2\right]G_{mn}\left(\omega\right)\right\}\\ &&-\mathrm{Re}\left\{ \frac{\pi}{2}\omega_{0}^{2}\left[f_0+\frac{1}{\omega_{0}}f_1+\frac{1}{\omega_{0}^{2}}f_2\right]G_{mn}\left(\omega_{0}\right)\right\}. \nonumber
\end{eqnarray}
Following Ref.~\onlinecite{Dzsotjan2011}, we now consider an integral along a closed contour in Fig.~\ref{fig:contour}
. We express the integral along the real semiaxis as a difference of the closed contour integral, integration along the imaginary semiaxis and integral along the curved contour $\mathcal{C}$ sufficiently distant for the integrated function to vanish: 
$\int_0^\infty\rightarrow \oint-\int_{i\infty}^0-\int_{\mathcal{C}}$. For the closed contour integral we make use of the residue theorem
\begin{equation*}
    \mathrm{Im}\left\{ \oint d\omega\frac{1}{2}\lim_{\epsilon\rightarrow0}\left(\frac{1}{\omega-\omega_{0}-i\epsilon}-\frac{1}{\omega+\omega_{0}-i\epsilon}\right)\omega^{2}\left[f_0+\frac{1}{\omega}f_1+\frac{1}{\omega^{2}}f_2\right]G_{mn}\left(\omega\right)\right\} = \pi\omega_{0}^{2}\left[f_0+\frac{1}{\omega_{0}}f_1+\frac{1}{\omega_{0}^{2}}f_2\right]G_{mn}^{\prime}\left(\omega_{0}\right).
\end{equation*}
Including integration along the imaginary axis and the contribution from $I_1$, we obtain
\begin{equation} \label{eqS:PV_replacement}
   I = \pi\omega_{0}^{2}\left[f_0+\frac{1}{\omega_{0}}f_1+\frac{1}{\omega_{0}^{2}}f_2\right]\mathrm{Re}G_{mn}\left(\omega_{0}\right)+\int_{0}^{\infty}d\kappa\kappa^{2}\left(\frac{\omega_{0}}{\kappa^{2}+\omega_{0}^{2}}\right)\mathrm{Re}\left\{\left(f_0+\frac{1}{i\kappa}f_1-\frac{1}{\kappa^{2}}f_2\right)G_{mn}\left(i\kappa\right)\right\}.
\end{equation}
\end{widetext}
Finally, we have replaced the principal value integral with an integration along the imaginary axis, where the integrated function in better behaved and more stable numerically. 

\section{Evaluation of Green's tensor at source's location in homogeneous media}\label{app:homogeneous}
We are interested to find the limit for $R\rightarrow 0$ of expression (\ref{eq:green_homogeneous}) of the main text, describing the homogeneous-medium Green's tensor. An off-diagonal element of the tensor is proportional to
\begin{equation}
G_{jk}(R,\omega) \sim \left(3-3ikR-k^2R^2 \right) e^{ikR}. 
\end{equation}
We focus on the case of atomic system's transition far-detuned from medium resonances, in which the refractive index is approximately real. Taylor-expanding the exponent around $R=0$ we find the imaginary part of the element above 

\begin{eqnarray}
\mathrm{Im}\, G_{jk}(R,\omega) &=& \left[\left(3-k^2R^2\right)\left(kR-\frac{k^3R^3}{3!}+\frac{k^5R^5}{5!}-...\right)\right.\nonumber\\
&&-\left.3kR\left(1-\frac{k^2R^2}{2!}+\frac{k^4R^4}{4!}-...\right)\right] \frac{R_jR_k}{4\pi k^2R^5}\nonumber\\
&=&\frac{k^3}{60\pi}R_jR_k +O(R^4).
\end{eqnarray}
Please note that all the lower-order terms vanish identically, i.e. the terms in the square bracket proportional to first and third powers of $R$. 
A similar calculation for diagonal terms leads to 
\begin{equation}
    \mathrm{Im}\, G_{jj}(R,\omega) = \frac{k}{6\pi}-\frac{k^3}{48\pi}R^2+\frac{k^3}{60\pi}R_jR_j+O(R^4).
\end{equation}

We now need to calculate Green's tensor's derivatives in Cartesian coordinates
\begin{eqnarray}
    \partial_m \mathrm{Im}\, G_{jk}(R\rightarrow 0,\omega) &=& \frac{k^3}{60\pi} \left(R_k \delta_{mj}+R_j \delta_{mk}\right),\\
    \partial_m \mathrm{Im}\, G_{jj}(R\rightarrow 0,\omega) &=& \partial_m \left(-\frac{k^3}{48\pi}\sum_k R_k^2+\frac{k^3}{60\pi}R_j^2\right) \nonumber\\
    &=& -\frac{k^3}{24\pi} R_m +\frac{k^3}{30\pi}R_j\delta_{mj}
\end{eqnarray}
Please note that the derivatives of diagonal and off-diagonal elements are comparable. Second-order derivatives are
\begin{eqnarray}
    \partial_n\partial_m \mathrm{Im}\, G_{jk}(R\rightarrow 0,\omega) &=& \frac{k^3}{60\pi} \left(\delta_{nk} \delta_{mj}+\delta_{nj}\delta_{mk}\right),\\
    \partial_n\partial_m \mathrm{Im}\, G_{jj}(R\rightarrow 0,\omega) &=& -\frac{k^3}{24\pi} \delta_{mn}+\frac{k^3}{30\pi}\delta_{mj}\delta_{nj}.
\end{eqnarray}

We now want to make a shift to derivatives over $\mathbf{r}$ and $\mathbf{r}^\prime$. We have 
$\frac{\partial f}{\partial r_k} = \sum_j\frac{\partial f}{\partial R_j}\frac{\partial R_j}{\partial r_k}$,
and similarly for $\mathbf{r}^\prime$ derivatives. In Cartesian coordinates
$R_j = r_j-r^\prime_j$, and $R=\sqrt{\sum_j R_j^2}$. Therefore
\begin{eqnarray}
    \frac{\partial}{\partial r_p} \mathrm{Im}\, G_{jk}(R\rightarrow 0,\omega) &=&
    \sum_m\frac{\partial}{\partial R_m} \mathrm{Im}\, G_{jk}(R\rightarrow 0,\omega) \delta_{mp} \nonumber \\
    &=&\frac{k^3}{60\pi} \left(R_k \delta_{pj}+R_j \delta_{pk}\right),
\end{eqnarray}
and
\begin{eqnarray}
\frac{\partial}{\partial r_p} \mathrm{Im}\, G_{jj}(R\rightarrow 0,\omega) &=&
\sum_m\frac{\partial}{\partial R_m} \mathrm{Im}\, G_{jj}(R\rightarrow 0,\omega) \delta_{mp}\nonumber\\
&=&-\frac{k^3}{24\pi} R_p +\frac{k^3}{30\pi}R_j\delta_{pj}.
\end{eqnarray}
Computation of second derivatives over primed variables leads to
\begin{equation}\label{eqS:2nd_derivative_offdiagonal}
    \frac{\partial}{\partial r^\prime_q}\frac{\partial}{\partial r_p} \mathrm{Im}\, G_{jk}(R\rightarrow 0,\omega) =
    -\left[\frac{k^3}{60\pi} \left(\delta_{kq} \delta_{pj}+\delta_{jq} \delta_{pk}\right)\right],
\end{equation}
and
\begin{equation}\label{eqS:2nd_derivative_diagonal}
    \frac{\partial}{\partial r^\prime_q}\frac{\partial}{\partial r_p} \mathrm{Im}\, G_{jj}(R\rightarrow 0,\omega) =\delta_{pq}\left(\frac{k^3}{15\pi}  -\frac{k^3}{30\pi}\delta_{jq}\right).
\end{equation}

\newpage
\begin{widetext}
\begin{center}
\textbf{\Large 
Supplementary Material for "Interaction of atomic systems with quantum vacuum beyond electric dipole approximation"}
\end{center}

\section*{Green's tensor of a pair of metallic nanospheres}

The following plots depict the imaginary part of the Green's tensor evaluated on the rectangular grid that is marked in Fig.~1(a) of the main text. The Green's tensor in each point of the grid is calculated from a source located in the same point. Thus the real part would always be infinite and we only show the imaginary part.

    \begin{figure}[b!]
        \centering
        \includegraphics[width=\linewidth]{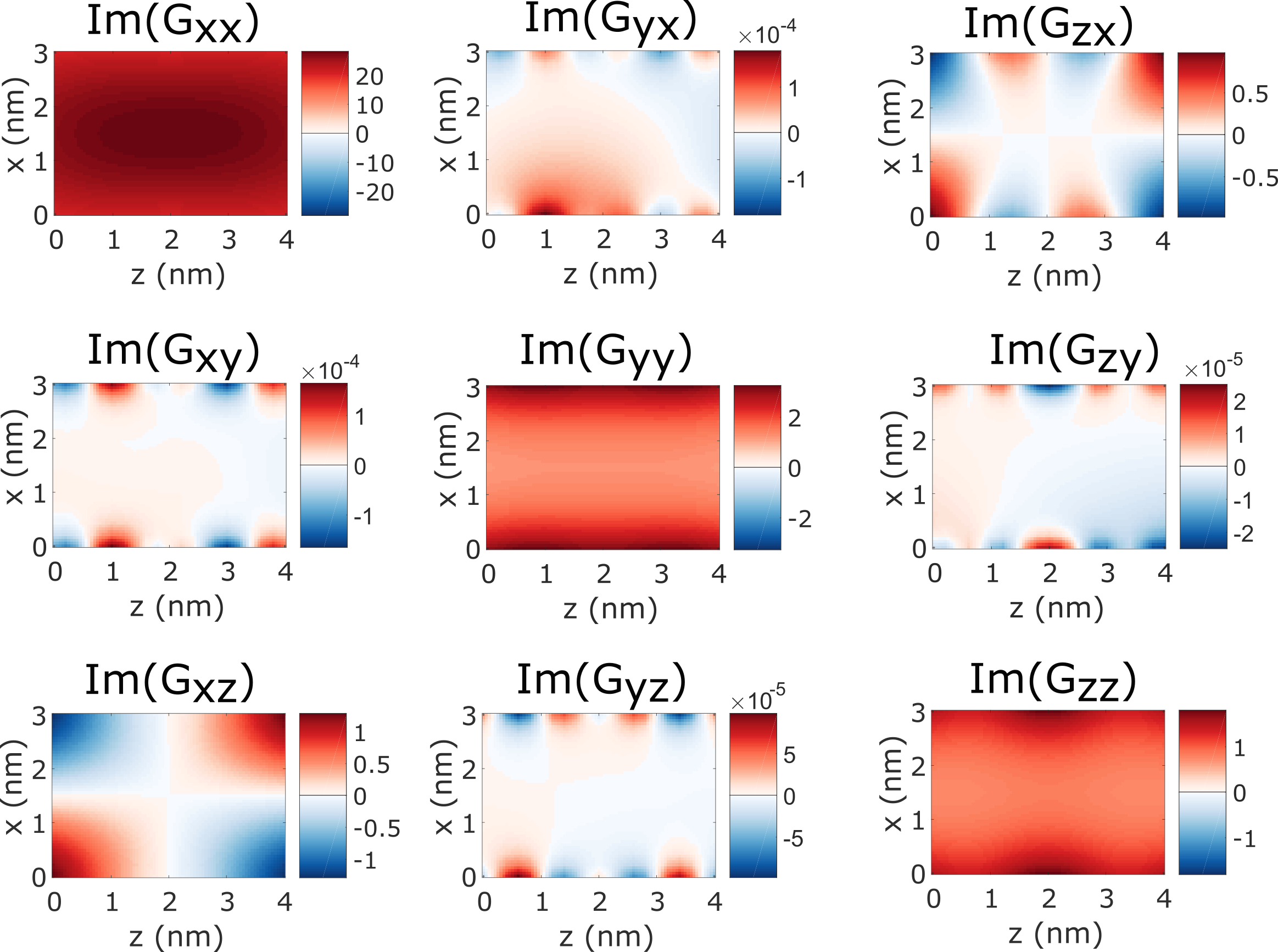}
        \caption[aa]{\small Green's tensor evaluated at the position of source (in nm$^{-1}$). } 
        \label{fig:tensor}
    \end{figure}
    
    \begin{figure}
        \centering
        \includegraphics[width=\linewidth]{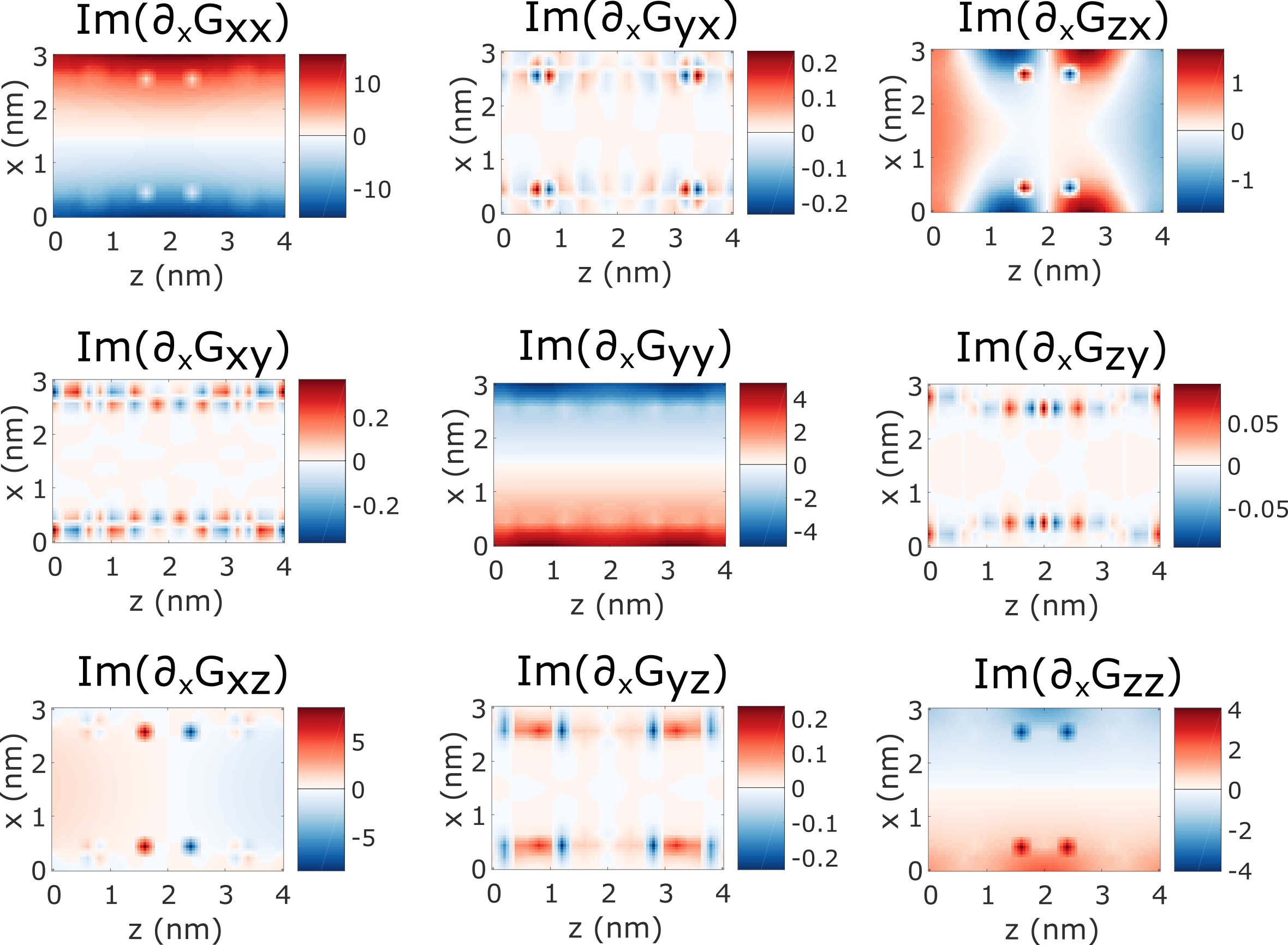}
        \caption[aa]{\small Green's tensor derivative along x direction evaluated at the position of source (in nm$^{-2}$). } 
    \end{figure}

    \begin{figure}
        \centering
        \includegraphics[width=\linewidth]{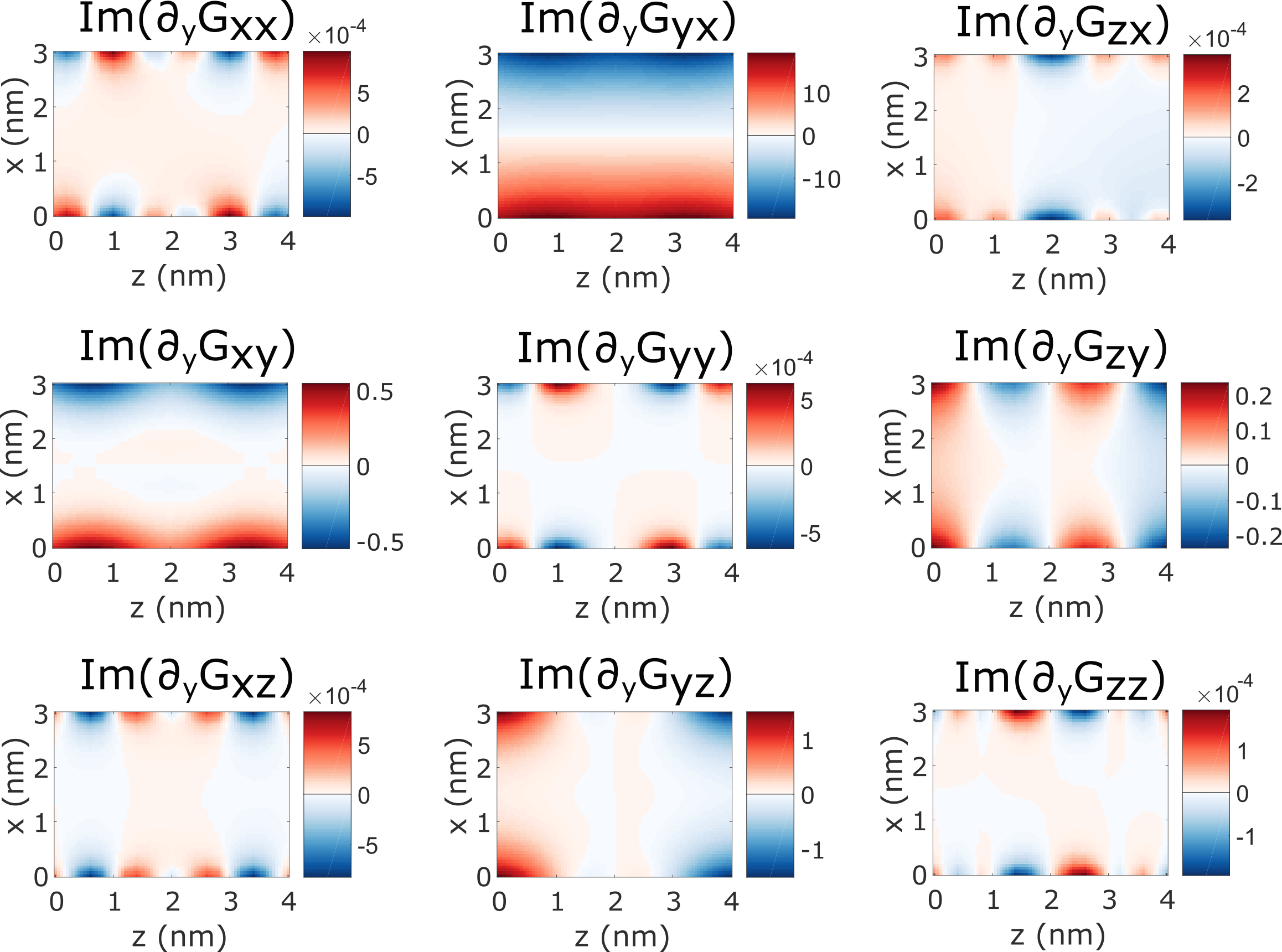}
        \caption[aa]{\small Green's tensor derivative along y direction evaluated at the position of source (in nm$^{-2}$).} 
    \end{figure}
    
    \begin{figure}
        \centering
        \includegraphics[width=\linewidth]{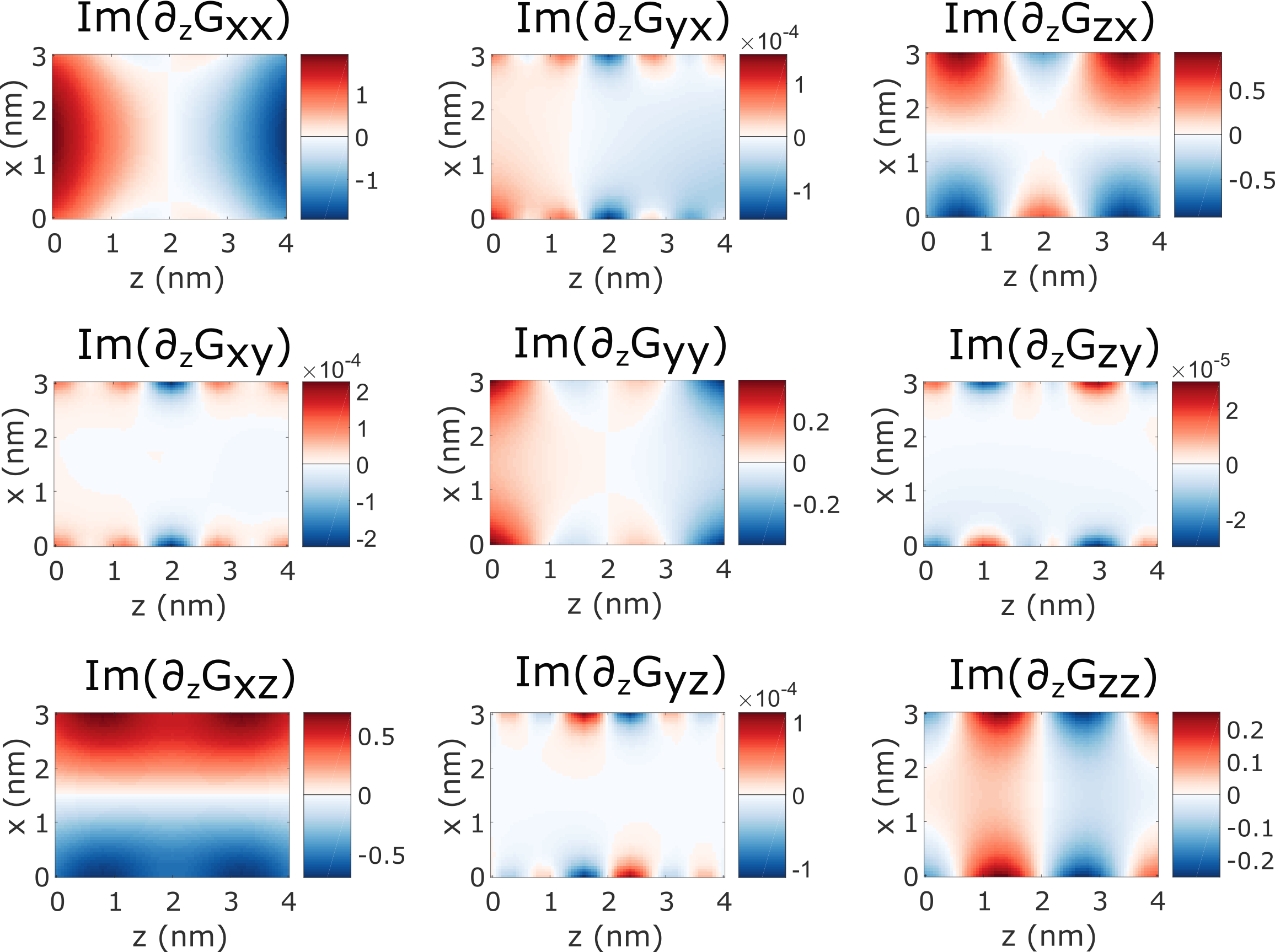}
        \caption[aa]{\small Green's tensor derivative along z direction evaluated at the position of source (in nm$^{-2}$). } 
    \end{figure}
    
    \begin{figure}
        \centering
        \includegraphics[width=\linewidth]{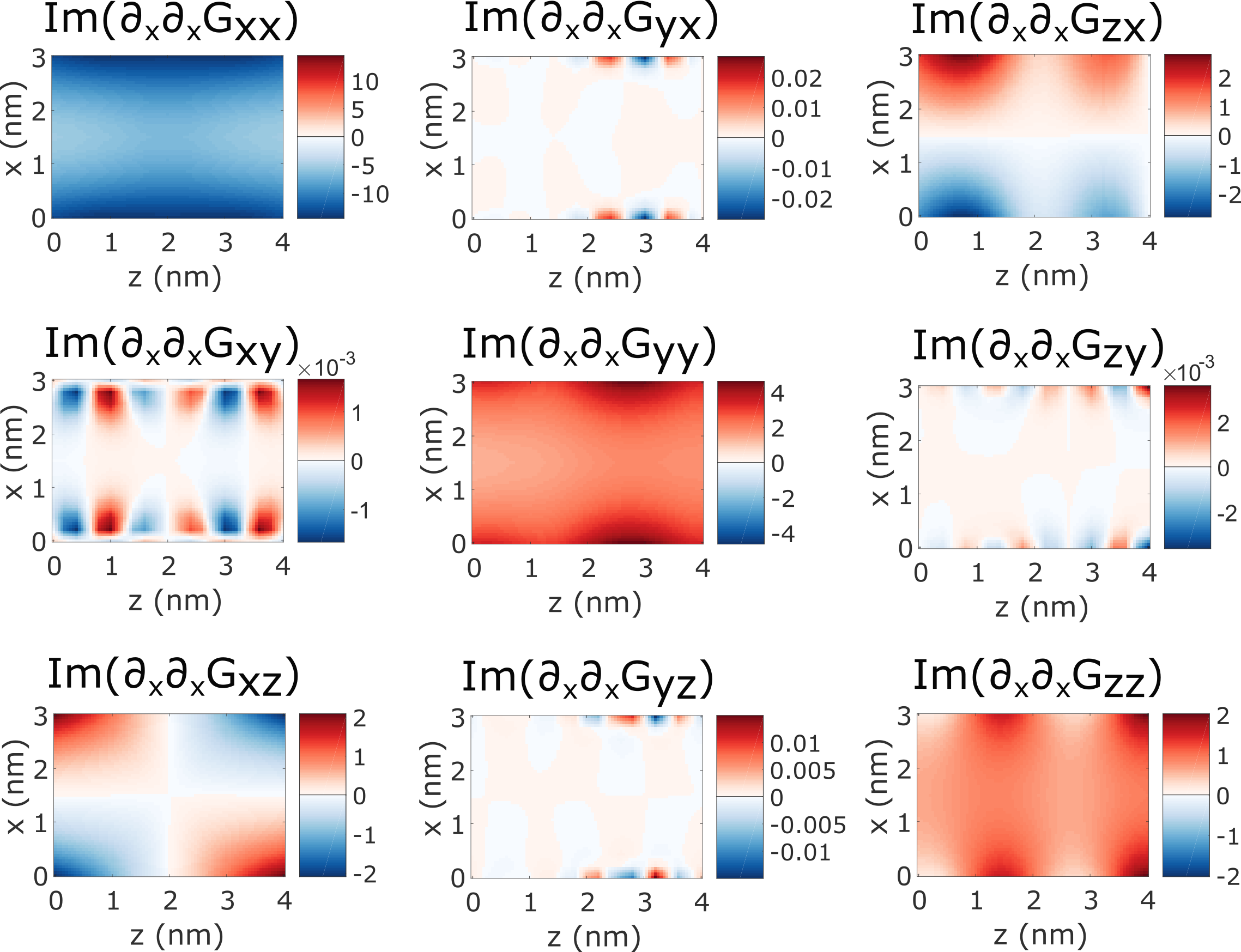}
        \caption[aa]{\small Green's tensor second derivative along x direction evaluated at the position of source (in nm$^{-3}$). } 
    \end{figure}

    \begin{figure}
        \centering
        \includegraphics[width=\linewidth]{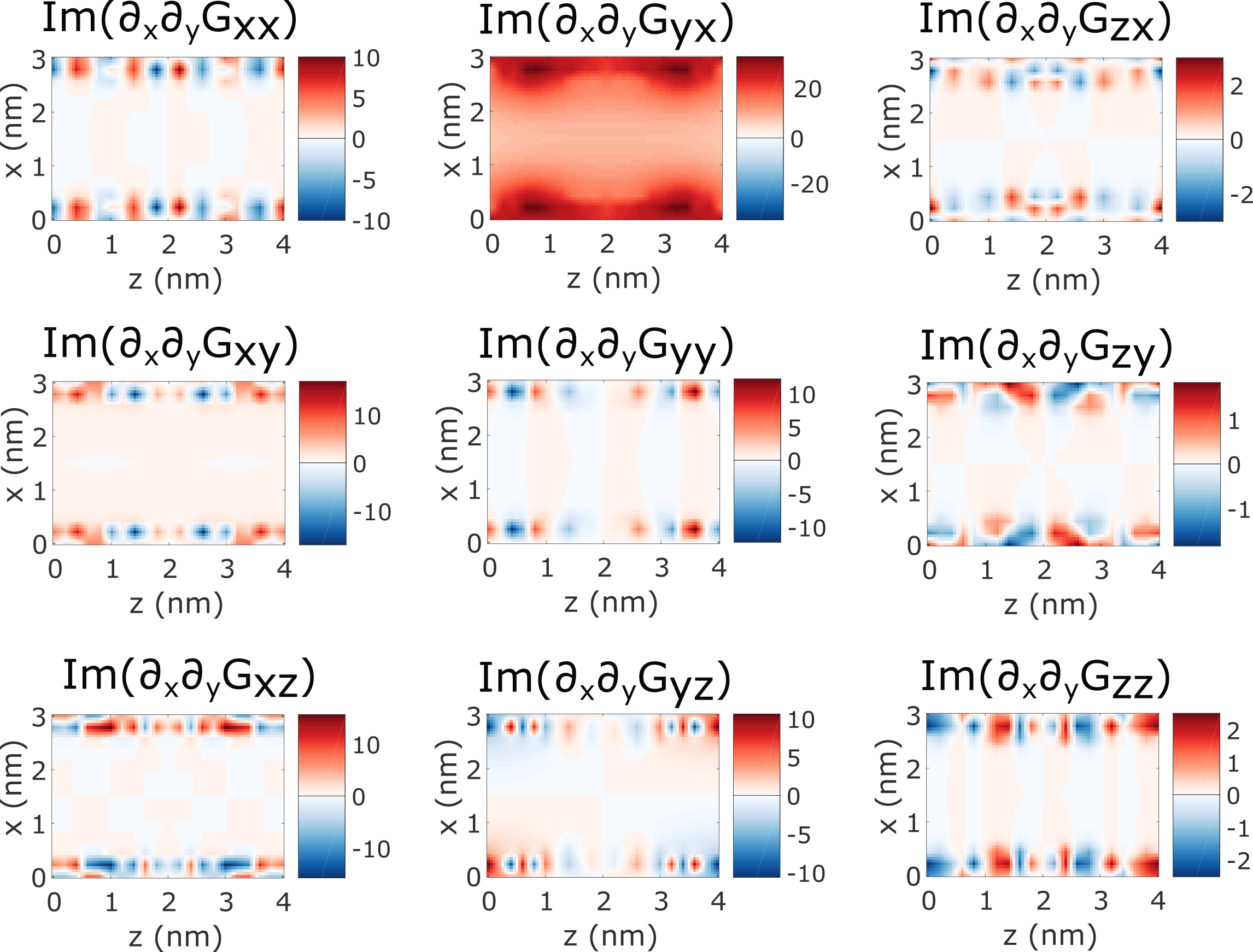}
        \caption[aa]{\small Green's tensor second derivative along x and along y direction evaluated at the position of source (in nm$^{-3}$). } 
    \end{figure}

    \begin{figure}
        \centering
        \includegraphics[width=\linewidth]{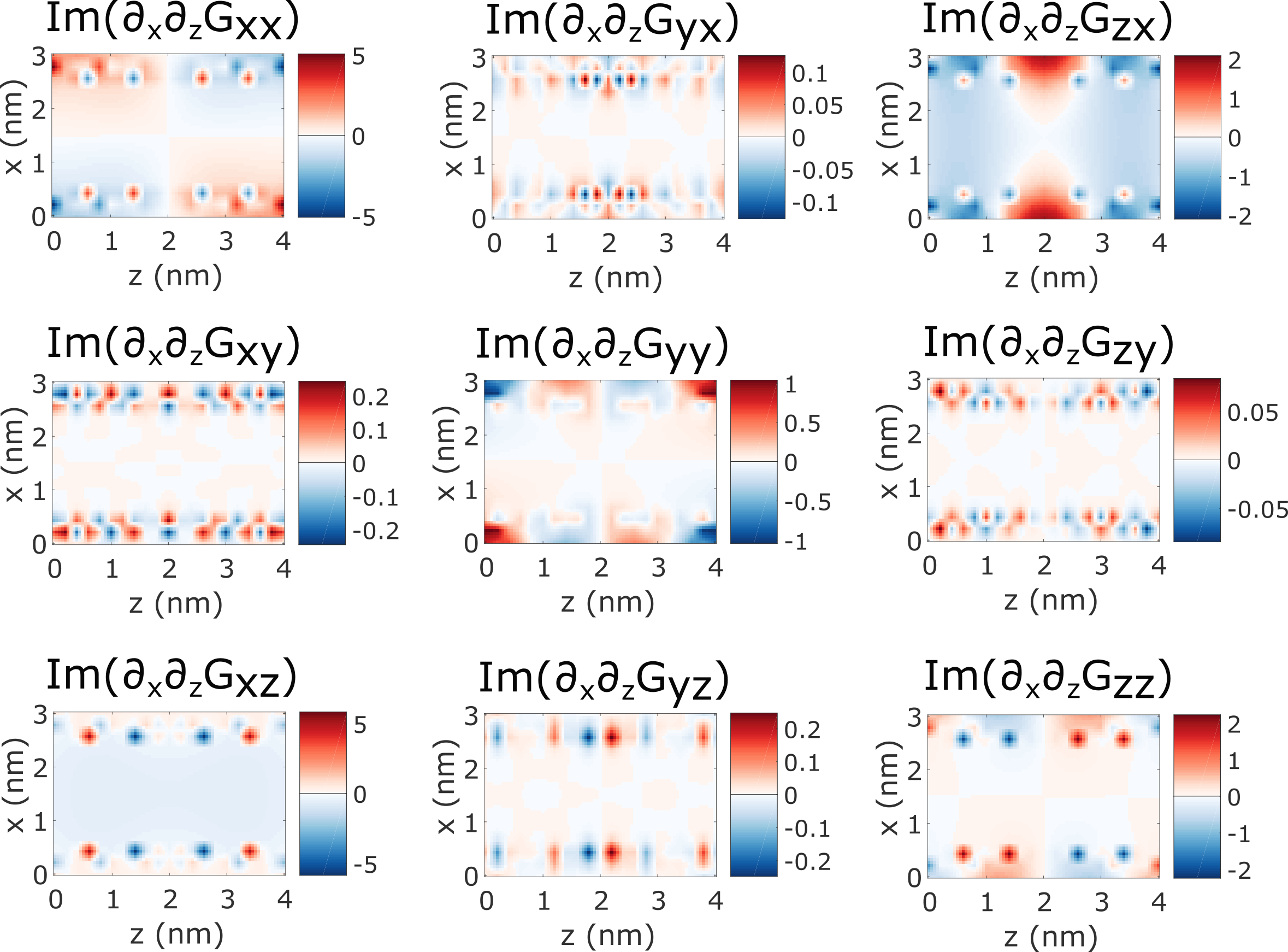}
        \caption[aa]{\small Green's tensor second derivative along x and along z direction evaluated at the position of source (in nm$^{-3}$). } 
    \end{figure}
    
    \begin{figure}
        \centering
        \includegraphics[width=\linewidth]{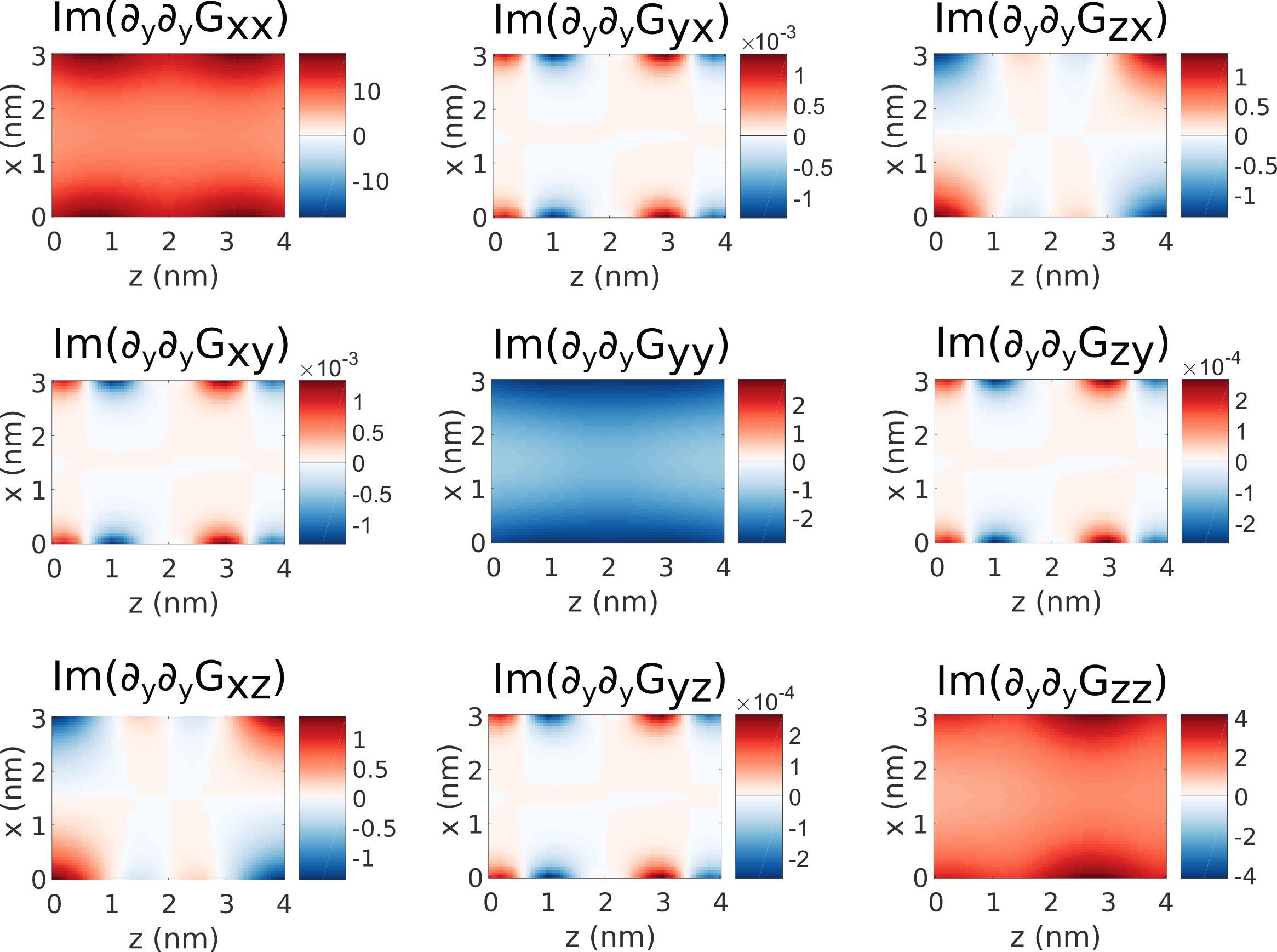}
        \caption[aa]{\small Green's tensor second derivative along y direction evaluated at the position of source (in nm$^{-3}$). } 
    \end{figure}
    
    \begin{figure}
        \centering
        \includegraphics[width=\linewidth]{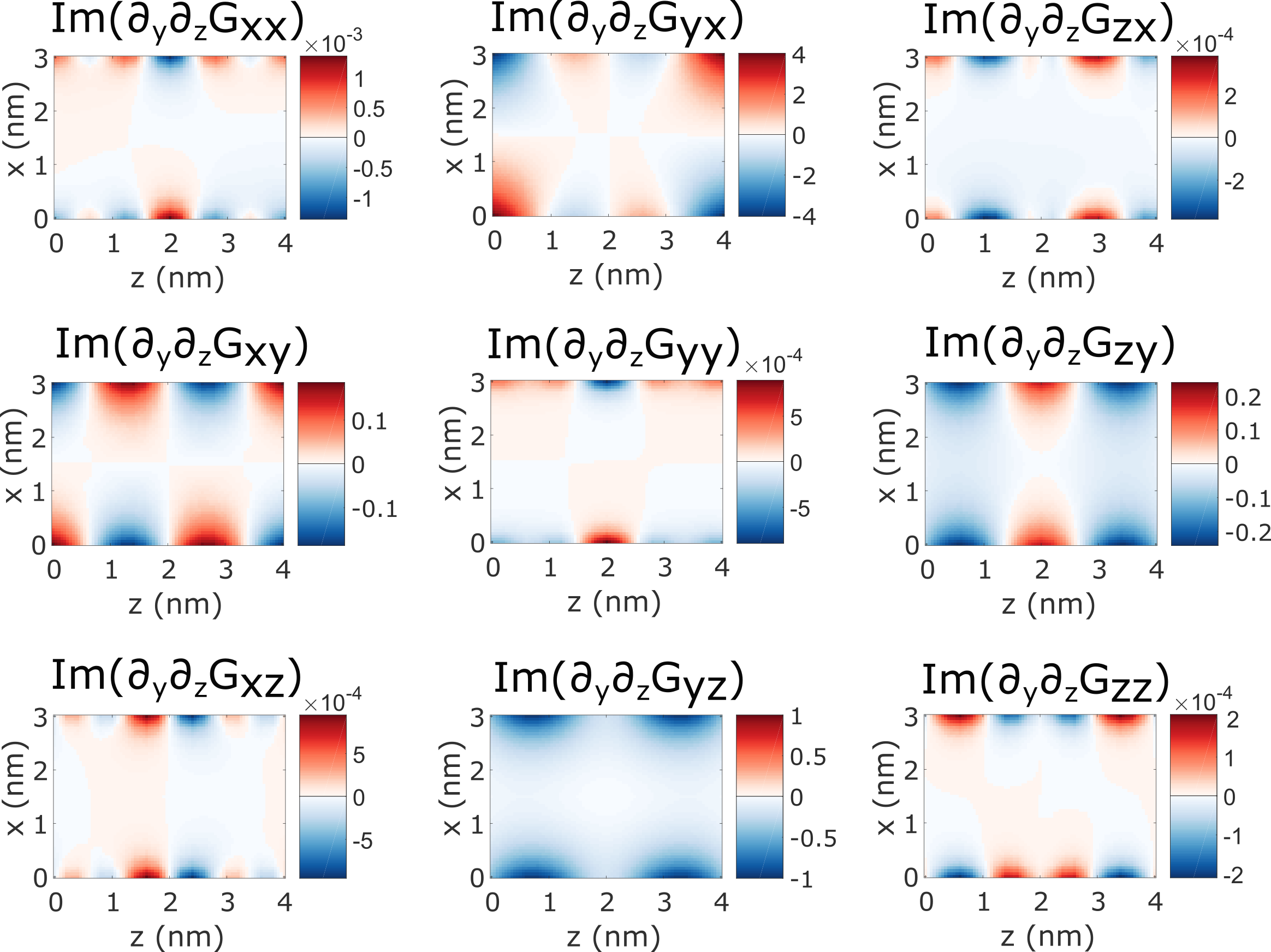}
        \caption[aa]{\small Green's tensor second derivative along y and along z direction evaluated at the position of source (in nm$^{-3}$). } 
    \end{figure}
    
    \begin{figure}
        \centering
        \includegraphics[width=\linewidth]{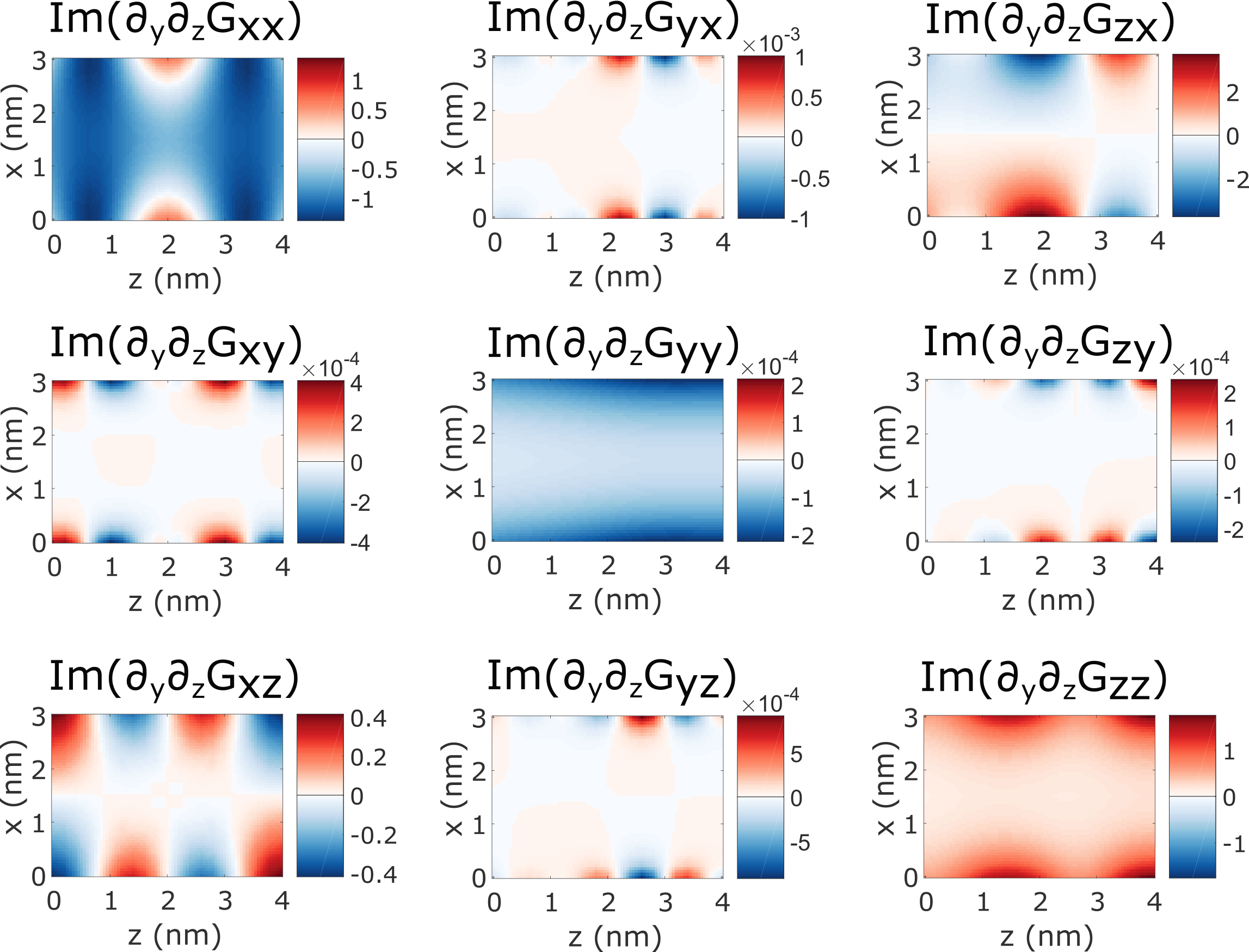}
        \caption[aa]{\small Green's tensor second derivative along z direction evaluated at the position of source (in nm$^{-3}$). } 
    \end{figure}
\end{widetext}    
\end{document}